\documentclass[10pt,english]{elsarticle}
\usepackage{lmodern}
\usepackage[T1]{fontenc}
\usepackage[latin9]{inputenc}
\usepackage{geometry}
\usepackage{setspace}
\usepackage{color}
\usepackage{babel}
\usepackage{array}
\usepackage{float}
\usepackage{bm}
\usepackage{amsbsy}
\usepackage{amstext}
\usepackage{graphicx}
\usepackage{setspace}
\usepackage[unicode=true,
 bookmarks=true,bookmarksnumbered=false,bookmarksopen=false,
 breaklinks=false,pdfborder={0 0 1},backref=false,colorlinks=false]
 {hyperref}

\makeatletter

\providecommand{\tabularnewline}{\\}

\usepackage{amssymb}
\usepackage{amsmath}
\usepackage{graphicx}
\usepackage{tensor}
\journal{Mathematics and Mechanics of Solids}
\date{}
\makeatother
\usepackage{times}
\begin{document}

\begin{frontmatter}{}

\title{Strain localization in reduced order asymptotic homogenization}

\author[rvt]{Harpreet Singh}

\ead{harpreet@iitgoa.ac.in}

\author[focal]{Puneet Mahajan}

\address[rvt]{School of Mechanical Sciences, Indian Institute of Technology Goa, Goa 403401, India }

\address[focal]{Department of Applied Mechanics, Indian Institute of Technology Delhi,
New Delhi 110001, India }
\begin{abstract} \begin{spacing}{1.1}\small
A reduced order asymptotic homogenization based multiscale technique
which can capture damage and inelastic effects in composite materials
is proposed. This technique is based on two scale homogenization procedure
where eigen strain representation accounts for the inelastic response
and the computational efforts are alleviated by reduction of order
technique. Macroscale stress is derived by calculating the influence
tensors from the analysis of representative volume element (RVE).
At microscale, the damage in the material is modeled using continuum
damage mechanics (CDM) based framework. To solve the problem of strain
localization a method of the alteration of stress-strain relation
of micro constituents based on the dissipated fracture energy in a
crack band is implemented. The issue of spurious post failure artificial
stiffness at macroscale is discussed and effect of increasing the
order to alleviate this problem is checked. Verification studies demonstrated
the proposed formulation predicts the macroscale response and also
captures the damage and plasticity induced inelastic strains. \end{spacing}
\end{abstract}
\begin{keyword} \begin{spacing}{1.1}\small
Multiscale Modeling \sep Asymptotic Expansion Homogenization \sep
Reduced Order Modeling \sep Damage \sep Plasticity\sep Strain Localization \end{spacing}
\end{keyword}

\end{frontmatter}{}

\section{Introduction}

Epoxy based fiber reinforced composites exhibit significant inelastic
deformation prior to failure of material. These inelastic deformations
also involve different damage mechanisms at micro constituent level.
These damages at microscale can be viewed in terms of degraded macroscale
stiffness. Continuum damage mechanics (CDM) provides a framework to
model this degradation of stiffness without using any explicit description
of the cracks. Macroscale modeling approach for fiber reinforced composite
material treats it as an anisotropic medium. An internal variable
is used to consider the damage and direction of cracks governs its
nature as a tensor. Similarly inelastic strains are calculated by
using an anisotropic potential function based plasticity formulation.
Phenomenological observations are directly implemented by adjusting
certain parameters. These kind of modeling approaches are easy to
implement but fail to achieve required level of accuracy. On the other
hand the second category is of multiscale models where the modeling
starts at the length scale of micro/meso level. By analyzing representative
section of composite microstructure the macroscale properties are
determined using various \textquotedbl{}homogenization\textquotedbl{}
techniques. These models have been found more accurate at the expense
of computational time and efforts involved.

A variety of homogenization techniques have been developed to calculate
the macroscopic properties from the analysis of microstructure. Asymptotic
expansion homogenization (AEH) is a rapidly developing field that
provides many benefits from computational cost point of view. Underlying
assumption of AEH is the periodicity of heterogeneity which can be
acceptable for long fiber composites. Historically emergence of AEH
started with pioneer work of \citet{bensoussan2011asymptotic} which
was followed by \citet{sanchez1980non}, \citet{guedes1990preprocessing},
\citet{hollister1992comparison} and \citet{terada2001class}. In
the presence of inelastic strains a two-scale AEH was firstly reported
by \citet{terada1995nonlinear}. \citet{fish1997computational} extended
AEH formulation for the prediction of inelastic strains by including
eigen strain and later on this formulation was further developed for
finite deformation plasticity of heterogeneous structure \citep{fish1999finite}.
Recently another eigen stain based AEH method was developed by \citet{zhang2015eigenstrain}
for polycrystalline materials and applied for the prediction of plastic
strain presented at macroscale. 

The damage effects were included in the classical AEH by introducing
an asymptotic expansion of damage variable by \citet{fish1999computational}.
By applying the principles of damage mechanics for each phase at microscale,
the macroscopic variation of the damage was calculated in their formulation.
Continuum damage mechanics based formulation used at microscale leads
to strain softening behavior of phase materials. Strain localization
associated with strain softening material leads to size effects. When
field variables are assumed to vary with $C^{0}$ continuity with
respect to the size of modeling zone, 'non-positive' material modeling
leads to non-objectivity. By using non-local formulations as proposed
by \citet{fish1999computational} this problem can be solved. This
however requires a known size of the influence volume. Various studies,
which were conducted in past \citep{fish2000multiscale,fish2005nonlocal,yuan2009multiple,oskay2010multiscale},
mostly use the non-local formulations to deal with the problem of
strain localization. In most of these formulations it is assumed that
the influence volume is confined in the domain of unit cell. Other
way of achieving the objectivity is by modifying the constitutive
law and making it depend on the size of zone of consideration \citep{bavzant1983crack,bavzant1989measurement,jirasek2012numerical}.
This approach is based on the appropriate adjustment of certain model
parameters that control the softening depending upon the size of partition.
This procedure makes the whole formulation local and the unknown variables
for each partition can be calculated by applying an averaging technique
on that local region. This kind of approach has been exploited in
past \citep{bavzant1984continuum,bavzant1989measurement} in various
finite element codes and also called by different names such as crack
band approach, mesh-adjusted softening modulus or fracture energy
tricks.

Apart from the challenge of spurious size/mesh dependent behavior
another one associated with numerical homogenization techniques is
the computational cost of solving RVE problem for the evaluation of
constitutive response at macroscale. One approach to alleviate this
issue is the use of reduced order technique for solving the RVE problem.
\citet{oskay2007eigendeformation} proposed a transformation field
analysis based reduced order formulation which accounts for interface
damage by using eigen-deformation based homogenization. The order
reduction at microscale also leads to coarse representation of inelastic
fields which causes artificial stiffness effects upon the failure
at the microscale. The false stress fields after complete failure
of the material corrupt the macroscale results and change the stress
distributions. 

In the present manuscript, the three areas of AEH based reduced order
modeling approach are addressed:
\begin{enumerate}
\item Major contribution of this manuscript in AEH research area can be
seen in terms of a combined formulation which discusses damage and
plasticity of the micro-constituents and captures the inelastic strains
caused by both.
\item An efficient way to solve ``\textbf{the problem of strain localization}''
is proposed for reduced order model which is based on the alteration
of constitutive laws of constituents. This alteration is governed
by the crack band theory which makes the fracture energy of each subdomain
independent of its size. 
\item ``\textbf{The problem of spurious residual stiffness}'' after the
complete failure at microscale and the effects of increasing the order
on the macroscale results are also discussed. 
\end{enumerate}
The manuscript is organized as follows. Section \ref{sec:Mathematical-framework-for}
explains the mathematical framework of reduced AEH homogenization
method and how the damage and plasticity induced effects at microscale
can be devised in the form of eigen strains. Finally this section
also discusses the method to obtain macroscale stress-strain response
using the homogenized properties. Section \ref{sec:Computational-Aspects}
explains the various computational aspects related to implementation
of reduced order AEH homogenization and crack band theory based solution
technique for strain localization problems. Section \ref{sec:Numerical-implementation-procedu}
demonstrates the numerical implementation procedure for proposed formulation
and also discusses the results of the simulations performed for a
RVE. At the end section \ref{sec:Macoscale-Simulation} shows the
\textcolor{black}{verification} study performed using homogenized properties calculated
from RVE simulations.

\section{Mathematical framework for multiscale modeling\label{sec:Mathematical-framework-for}}

Mathematical framework for the homogenization starts with the assumption
that the under-consideration material has a microstructure comprised
of two or more phases and shape and orientation of these phases leads
to a periodic three dimensional network of a representative element
of volume. For fiber-reinforced composites, a RVE is generated, which
represents the basic fiber-matrix structure, at each Gauss point in
the finite elements constituting the macroscale domain. Next macroscale
field functions are represented as asymptotic expansions and based
on those a boundary value problem is formulated to obtain various
influence functions. These functions collectively represent the behavior
of the material at next scale. For two-scale formulation of fiber-reinforced
composites, the next scale is regarded as scale of lamina. The homogenized
properties and stress-strain response at lamina-scale are calculated
by using the influence functions. Section \ref{sec:Mathematical-framework-for} explains the procedure
to determine the homogenized properties and macroscale stress using
reduced order modeling approach. 

\subsection{Definition of scales}

Let us consider the composite material consisting of a domain, $\Gamma$
and \textcolor{black}{at each material point, there exists a periodically repeating microstructure.} This periodic subdomain is also known as statistically
homogeneous RVE, denoted as $\Theta$, and consists of multiple
materials. Let $\boldsymbol{x}=(x_{1},x_{2},x_{3})$ be a position
vector defined at macroscale for domain, $\Gamma$. Similarly $\boldsymbol{y}=(y_{1},y_{2},y_{3})$
is the position vector defined at microscale for subdomain, $\Theta$. Coordinates
represented by the position vector, $\boldsymbol{y}$ may not represent
the actual size of the microstructure and denote the artificially
scaled up version of the actual microstructure. A positive \textcolor{black}{dimensionless} number,
$\xi$ denotes the scale factor \textcolor{black}{($\xi<<1$)} which defines the ratio of \textcolor{black}{the size of microstructure measured} in macroscale
$\boldsymbol{x}$ dimension to microscale $\boldsymbol{y}$ dimension.
Due to the periodic distribution of heterogeneities, the response
fluctuations are also periodic in nature. \textcolor{black}{Any response function, exists for the domain, depends upon multiple scales due to existence of this periodic microstructure}. Let a vector, $\hat{\boldsymbol{y}}$
denotes the basic period of RVE then microscale periodic response
function, $f$ has the following property:

\begin{equation}
f(\boldsymbol{x},\boldsymbol{y})=f(\boldsymbol{x},\boldsymbol{y}+k\hat{\boldsymbol{y}})\label{eq:1}
\end{equation}
where $k=(k_{1},k_{2},k_{3})$ and $k_{1}$, $k_{2}$ and $k_{3}$
are the integers. The response function, $f$ which not only depends
upon the macroscale coordinates, $\boldsymbol{x}$ but also on microscale
dimension, $\boldsymbol{y}$, can be denoted as 

\begin{equation}
f^{\xi}(\boldsymbol{x})=f(\boldsymbol{x},\boldsymbol{y})\label{eq:2}
\end{equation}
\textcolor{black}{ Superscript '$\xi$' indicates the dependence of $f$ on microstructural heterogeneity.}
The macroscopic spatial derivatives for $f^{\xi}(\boldsymbol{x})$ 
can be calculated as 

\begin{equation}
f_{,x_{i}}^{\xi}(\boldsymbol{x})=f_{,x_{i}}(\boldsymbol{x},\boldsymbol{y})+\frac{1}{\xi}f_{,y_{i}}(\boldsymbol{x},\boldsymbol{y})\label{eq:3}
\end{equation}

\subsection{Macroscale problem definition}

Considering the static equilibrium and assuming small deformations,
the governing differential equations for macroscale domain can be
written as

\begin{eqnarray}
\frac{\partial\sigma_{ij}^{\xi}(\boldsymbol{x})}{\partial x_{j}}+b_{i}^{\xi}(\boldsymbol{x}) & = & 0\label{eq:4}\\
\sigma_{ij}^{\xi}(\boldsymbol{x}) & = & L_{ijkl}^{\xi}(\boldsymbol{x})(\epsilon_{kl}^{\xi}(\boldsymbol{x})-\mu_{kl}^{\xi}(\boldsymbol{x}))\label{eq:5}\\
\epsilon_{ij}^{\xi}(\boldsymbol{x}) & = & u_{\textcolor{black}{(i,j)}}^{\xi}(\boldsymbol{x})=\frac{1}{2}\left(\frac{\partial u_{i}^{\xi}(\boldsymbol{x})}{\partial x_{j}}+\frac{\partial u_{j}^{\xi}(\boldsymbol{x})}{\partial x_{i}}\right)\qquad\text{such that}\quad i,j,k,l\in\{1,2,3\}\label{eq:6}
\end{eqnarray}
where $\sigma_{ij}^{\xi}$ and $\epsilon_{ij}^{\xi}$ are components
of Cauchy stress and strain. $\mu_{ij}^{\xi}$ is the eigen strain
which depends upon loading history. $b_{i}^{\xi}$ and $u_{i}^{\xi}$
are the components of body force and displacement. $L_{ijkl}^{\xi}$
is elastic modulus tensor. \textcolor{black}{Subscript '($\boldsymbol{\cdot}$)' denotes the symmetric part of the quantity}. The periodic microstructure
leads to the fact that $L_{ijkl}^{\xi}$ is only function of microscale
dimension, $\boldsymbol{y}$ and do not depend on the macroscale coordinates,
$\boldsymbol{x}$. Therefore, the elasticity tensor can be denoted
as

\textcolor{black}{
\begin{equation}
L_{ijkl}^{\xi}(\boldsymbol{x})=L_{ijkl}(\boldsymbol{y})
\end{equation}
}

The surface boundary, $\partial\Gamma$ of the $\Gamma$-domain may
be divided into two parts, $\partial\Gamma_{u}$ and $\partial\Gamma_{t}$
such that $\partial\Gamma_{u}\cup\partial\Gamma_{t}=\partial\Gamma$
and $\partial\Gamma_{u}\cap\partial\Gamma_{t}=\emptyset$. The following
boundary conditions are applicable:

\begin{eqnarray}
u_{i}^{\xi}(\boldsymbol{x}) & = & \hat{u}_{i}(\boldsymbol{x})\qquad\forall\boldsymbol{x}\in\partial\Gamma_{u}\label{eq:7}\\
\sigma_{ij}^{\xi}(\boldsymbol{x})n_{j} & = & \hat{p}(\boldsymbol{x})\qquad\forall\boldsymbol{x}\in\partial\Gamma_{t}\label{eq:8}
\end{eqnarray}
where $\hat{u}_{i}(\boldsymbol{x})$ and $\hat{p}_{i}(\boldsymbol{x})$
are prescribed displacement and traction on $\partial\Gamma_{u}$
and $\partial\Gamma_{t}$ respectively. $\boldsymbol{n}$ is the unit
normal to the surface $\partial\Gamma_{t}$.

\begin{figure}[H]
\begin{centering}
\includegraphics[width=0.85\textwidth]{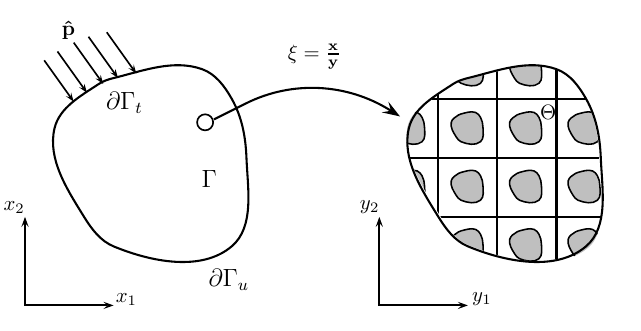}
\par\end{centering}
\caption{Mapping of macroscale ($\Gamma$) domain under various loading conditions
to microscale ($\Theta$) domain.\label{fig:1}}
\end{figure}

\subsection{Asymptotic expansion homogenization}

Two scale $(\Gamma\times\Theta)$ asymptotic expansion homogenization
(AEH) procedure is adopted to get the macroscopic response using the
microscopic details and properties. The advantage of AEH methodology
is that it enables the macroscopic properties to be calculated in
terms of some characteristic microscale functions which are termed
as influence functions. Using this method the response functions at
the global or macroscale can be represented as

\begin{equation}
u_{i}^{\xi}(\boldsymbol{x})=u_{i}^{(0)}(\boldsymbol{x},\boldsymbol{y})+\xi u_{i}^{(1)}(\boldsymbol{x},\boldsymbol{y})+\xi^{2}u_{i}^{(2)}(\boldsymbol{x},\boldsymbol{y})+\cdots\label{eq:10}
\end{equation}
where $u_{i}^{(s-1)}(\boldsymbol{x},\boldsymbol{y})$ with $s\in\mathbb{N}$
are the periodic functions in microscale. Now using Eq.(\ref{eq:10})
and Eq.(\ref{eq:3}), Eq.(\ref{eq:6}) can be expressed as following:

\begin{equation}
\epsilon_{ij}^{\xi}(\boldsymbol{x})=\xi^{-1}\epsilon_{ij}^{(-1)}(\boldsymbol{x},\boldsymbol{y})+\xi^{0}\epsilon_{ij}^{(0)}(\boldsymbol{x},\boldsymbol{y})+\xi^{1}\epsilon_{ij}^{(1)}(\boldsymbol{x},\boldsymbol{y})+\cdots\label{eq:11}
\end{equation}
where 

{\small{}
\begin{eqnarray}
\epsilon_{ij}^{(-1)} & = & \frac{\partial u_{i}^{(0)}}{\partial y_{j}}\label{eq:12}\\
\epsilon_{ij}^{(s)} & = & \frac{1}{2}\left(\frac{\partial u_{i}^{(s)}}{\partial x_{j}}+\frac{\partial u_{j}^{(s)}}{\partial x_{i}}+\frac{\partial u_{i}^{(s+1)}}{\partial y_{j}}+\frac{\partial u_{j}^{(s+1)}}{\partial y_{i}}\right)\qquad\forall s\in\{0,1,2,\cdots\}\label{eq:13}
\end{eqnarray}
}{\small \par}

The eigen strain can also be expressed as

\begin{equation}
\mu_{ij}^{\xi}(\mathbf{x})=\mu_{ij}^{(0)}(\boldsymbol{x},\boldsymbol{y})+\xi\mu_{ij}^{(1)}(\boldsymbol{x},\boldsymbol{y})+\xi^{2}\mu_{ij}^{(2)}(\boldsymbol{x},\boldsymbol{y})+\cdots\label{eq:14}
\end{equation}

Inserting Eq.(\ref{eq:11}) into Eq.(\ref{eq:5}) gives

\begin{equation}
\sigma_{ij}^{\xi}(\mathbf{x})=\xi^{-1}\sigma_{ij}^{(-1)}(\boldsymbol{x},\boldsymbol{y})+\xi^{0}\sigma_{ij}^{(0)}(\boldsymbol{x},\boldsymbol{y})+\xi^{1}\sigma_{ij}^{(1)}(\boldsymbol{x},\boldsymbol{y})+\cdots\label{eq:15}
\end{equation}
where

\begin{eqnarray}
\sigma_{ij}^{(-1)}(\boldsymbol{x},\boldsymbol{y}) & = & L_{ijkl}(\boldsymbol{y})\epsilon_{kl}^{(-1)}(\boldsymbol{x},\boldsymbol{y})\label{eq:16}\\
\sigma_{ij}^{(s)}(\boldsymbol{x},\boldsymbol{y}) & = & L_{ijkl}(\boldsymbol{y})\left(\epsilon_{kl}^{(s)}(\boldsymbol{x},\boldsymbol{y})-\mu_{kl}^{(s)}(\boldsymbol{x},\boldsymbol{y})\right)\qquad\forall s\in\{0,1,2,\cdots\}\label{eq:17}
\end{eqnarray}

\subsection{Microscale problem definition}

Using Eq.(\ref{eq:15}), Eq.(\ref{eq:4}) can be finally written as 

\begin{equation}
\frac{1}{\xi^{2}}\frac{\partial\sigma_{ij}^{(-1)}}{\partial y_{j}}+\frac{1}{\xi}\left(\frac{\partial\sigma_{ij}^{(0)}}{\partial y_{j}}+\frac{\partial\sigma_{ij}^{(-1)}}{\partial x_{j}}\right)+\frac{1}{\xi^{0}}\left(\frac{\partial\sigma_{ij}^{(0)}}{\partial x_{j}}+\frac{\partial\sigma_{ij}^{(1)}}{\partial y_{j}}+b_{i}\right)+\cdots=0\label{eq:18}
\end{equation}

In order to make Eq.(\ref{eq:18}) valid for any value of $\xi$,
following equations are expressed:

\begin{eqnarray}
 & \mathcal{O}(\xi^{-2})\qquad & \frac{\partial\sigma_{ij}^{(-1)}}{\partial y_{j}}=0\label{eq:19}\\
 & \mathcal{O}(\xi^{-1})\qquad & \frac{\partial\sigma_{ij}^{(-1)}}{\partial x_{j}}+\frac{\partial\sigma_{ij}^{(0)}}{\partial y_{j}}=0\label{eq:20}\\
 & \mathcal{O}(\xi^{0})\qquad & \frac{\partial\sigma_{ij}^{(0)}}{\partial x_{j}}+\frac{\partial\sigma_{ij}^{(1)}}{\partial y_{j}}+b_{i}=0\label{eq:21}\\
 & \mathcal{O}(\xi^{s})\qquad & \frac{\partial\sigma_{ij}^{(s)}}{\partial x_{j}}+\frac{\partial\sigma_{ij}^{(s+1)}}{\partial y_{j}}=0\qquad\forall s\in\{1,2,\cdots\}\label{eq:22}
\end{eqnarray}

Due to periodic nature of the microstructure $\mathcal{O}(\xi^{-2})$
can be deduced to $u_{i}^{(0)}=u_{i}^{(0)}(\boldsymbol{x})$ and $\sigma_{ij}^{(-1)}(\boldsymbol{x},\boldsymbol{y})=0$.
$\mathcal{O}(\xi^{-1})$ term (Eq.(\ref{eq:20})) can be expressed
as 

\begin{equation}
\left\{ L_{ijkl}(\boldsymbol{y})\left(\frac{\partial u_{k}^{(0)}}{\partial x_{l}}+\frac{\partial u_{k}^{(1)}}{\partial y_{l}}-\mu_{kl}^{(0)}(\boldsymbol{x},\boldsymbol{y})\right)\right\} _{,y_{j}}=0\label{eq:23}
\end{equation}

Using the fact that the term $\frac{\partial u_{k}^{(0)}}{\partial x_{l}}$
is a constant w.r.t. operator '$,y_{j}$', the first order deformation
can be represented in terms of macroscale strain and eigen strain
as

\begin{equation}
u_{i}^{(1)}(\boldsymbol{x},\boldsymbol{y})=H_{i}^{kl}(\boldsymbol{y})u_{k,x_{l}}^{(0)}(\boldsymbol{x})+\int_{\Theta}\chi_{i}^{kl}(\boldsymbol{y},\tilde{\boldsymbol{y}})\mu_{kl}^{(0)}(\boldsymbol{x},\tilde{\boldsymbol{y}})\,d\tilde{\Theta}\label{eq:24}
\end{equation}
where $H_{i}^{kl}(\boldsymbol{y})$ and $\chi_{i}^{kl}(\boldsymbol{y},\tilde{\boldsymbol{y}})$
are \textcolor{black}{strain and eigen strain influence functions }that
relate the first order displacement to macroscale strain and eigen
strain respectively. Finally Eq.(\ref{eq:23}) can be expressed as

\begin{equation}
\left\{ L_{ijkl}(\boldsymbol{y})\left(E_{klmn}(\boldsymbol{y})\frac{\partial u_{m}^{(0)}}{\partial x_{n}}+\int_{\Theta}S_{klmn}(\boldsymbol{y},\tilde{\boldsymbol{y}})\mu_{kl}^{(0)}(\boldsymbol{x},\tilde{\boldsymbol{y}})\,d{\tilde{\Theta}}\right)\right\} _{,y_{j}}=0\label{eq:25}
\end{equation}
where $I_{klmn}=\frac{1}{2}\{\delta_{km}\delta_{ln}+\delta_{lm}\delta_{kn}\}$
is a fourth order identity tensor and

\begin{eqnarray}
E_{klmn}(\boldsymbol{y}) & = & \left[I_{klmn}+H_{k,y_{l}}^{mn}(\boldsymbol{y})\right]\label{eq:26}\\
{S_{klmn}}(\boldsymbol{y},\tilde{\boldsymbol{y}}) & = & {\chi}_{k,y_{l}}^{mn}(\boldsymbol{y},\tilde{\boldsymbol{y}})-I_{klmn}\delta(\boldsymbol{y}-\tilde{\boldsymbol{y}})\label{eq:27}
\end{eqnarray}
Using Eq.(\ref{eq:13}), zero order strain is written as

\begin{equation}
\epsilon_{ij}^{(0)}(\boldsymbol{x},\boldsymbol{y})=E_{ijkl}(\boldsymbol{y})u_{k,x_{l}}^{(0)}(\boldsymbol{x})+\int_{\Theta}S_{klmn}(\boldsymbol{y},\tilde{\boldsymbol{y}})\mu_{kl}^{(0)}(\boldsymbol{x},\tilde{\boldsymbol{y}})\,d{\tilde{\Theta}}\label{eq:28}
\end{equation}

\subsection{Reduction of order modeling}

Eq.(\ref{eq:25}) can be solved numerically discretizing the domain
into infinite set of points which may not be computationally economical.
The computational cost can be made affordable by subdividing the domain
into finite number of sets which results in reduction of the order
of microscale domain. So the weighted average of response functions
over these sub-domains represents the the response of RVE. 

\begin{figure}[H]
\begin{centering}
\includegraphics[width=0.7\textwidth]{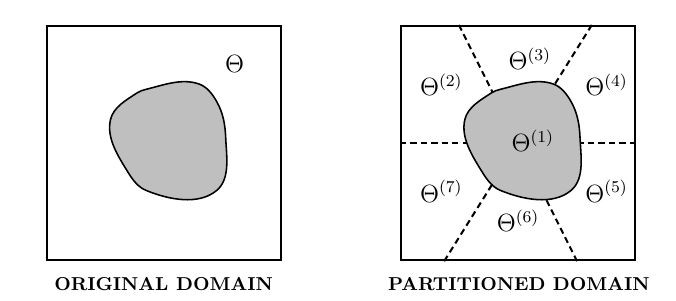}
\par\end{centering}
\caption{Partition of $\Theta$ domain into number of sub-domains with single
partition for fiber and six partitions for matrix.\label{fig:2}}
\end{figure}

Based on this, the eigen strain, $\bm{\mu}^{(0)}$ can be defined
as

\begin{equation}
\mu_{ij}^{(0)}(\boldsymbol{x},\boldsymbol{y})=\sum_{\beta=1}^{M}N^{(\beta)}(\boldsymbol{y})\mu_{ij}^{(\beta)}(\boldsymbol{x})\label{eq:29}
\end{equation}
where $\beta$ designates a particular set and $M$ denotes the finite
number of sets which are used to represents the response of microstructure.
Fig. \ref{fig:2} shows RVE sub-domain discretization for $M$ = 7.
As shown in Fig. \ref{fig:2}, it is generally preferred to use common
surface boundaries of various phases as full or partial sub-domain
boundaries. The eigen strains may be discontinuous across these sub-domains
and have $\mathcal{C}^{-1}$ continuity. The functions $N^{(\beta)}$
are also assumed to follow the condition of partition of unity ($\sum_{\beta=1}^{M}N^{(\beta)}(\boldsymbol{y})=1$).
The choice regarding the shape functions is further simplified by
using the following:

\begin{equation}
N^{(\beta)}(\boldsymbol{y})=\begin{cases}
1 & \boldsymbol{y}\in\Theta^{(\beta)}\\
0 & \boldsymbol{y}\notin\Theta^{(\beta)}
\end{cases}\label{eq:30}
\end{equation}

The eigen strain for a particular partition, $\beta$ can be calculated
as

\begin{equation}
\mu_{ij}^{(\beta)}(\boldsymbol{x})=\int_{\Theta}\bar{\varphi}^{(\beta)}\mu_{ij}^{(0)}(\boldsymbol{x},\boldsymbol{y})\,d\Theta\label{eq:31}
\end{equation}
where $\bar{\varphi}^{(\beta)}$ is associated weight function for
the partition. $\bar{\varphi}^{(\beta)}$ may be treated as localization
limiter which changes the nature of eigen strain from local to nonlocal.
Further $\bar{\varphi}^{(\beta)}$ can be calculated as

\begin{equation}
\bar{\varphi}^{(\beta)}=\frac{\varphi(\boldsymbol{y}-\mathbf{\zeta})}{\int\varphi(\boldsymbol{y}-\mathbf{\zeta})\,d\zeta}\label{eq:32}
\end{equation}

This $\varphi(\boldsymbol{y}-\mathbf{\zeta})$ can be assumed as a
distribution function such as Normal or Gaussian which involves an
additional parameter, $\mathbf{\zeta}$ to define the interaction
distance. Calculation of this parameter is itself ambiguous and difficult
to determine. 

Similar to eigen strain calculation the fine scale strain for a particular
partition, $\beta$ can be calculated as

\begin{equation}
{\boldsymbol{\epsilon}^{(\beta)}}(\boldsymbol{x})=\mathbf{E}^{(\beta)}:\frac{\partial\boldsymbol{u}^{(0)}}{\partial\boldsymbol{x}}+\sum_{\alpha=1}^{M}\mathbf{S}^{(\alpha\beta)}:\boldsymbol{\mu}^{(\alpha)}(\boldsymbol{x})\label{eq:38}
\end{equation}
where 

\begin{equation}
\mathbf{E}^{(\beta)}=\int_{\Theta}\bar{\varphi}^{(\beta)}E_{ijkl}(\boldsymbol{y})\,d\Theta\label{eq:39}
\end{equation}
and

\begin{equation}
\mathbf{S}^{(\alpha\beta)}=\int_{\Theta}\bar{\varphi}^{(\beta)}S_{ijkl}^{(\alpha)}(\boldsymbol{y})\,d\Theta\label{eq:40}
\end{equation}
\textcolor{black}{$\mathbf{E}^{(\beta)}$ and $\mathbf{S}^{(\alpha\beta)}$
are called as elastic and phase damage coefficient tensors respectively.
}Macroscopic stress can be calculated by taking the average over the
microscale domain as:

\begin{equation}
\bar{\sigma}_{ij}^{\text{}}(\boldsymbol{x})=\frac{1}{\mid\Theta\mid}\int_{\Theta}\sigma_{ij}^{(0)}(\boldsymbol{x},\boldsymbol{y})\,d\Theta\label{eq:33}
\end{equation}

Using Eq.(\ref{eq:28}), Eq.(\ref{eq:33}) can be expressed as 

{\small{}
\begin{equation}
\bar{\sigma}_{ij}^{\text{}}(\boldsymbol{x})=\left[\frac{1}{\mid\Theta\mid}\int_{\Theta}L_{ijkl}(\boldsymbol{y})E_{klpq}(\boldsymbol{y})\,d\Theta\right]u_{p,x_{q}}^{(0)}(\boldsymbol{x})+\sum_{\alpha=1}^{M}\left[\frac{1}{\mid\Theta\mid}\int_{\Theta}L_{ijkl}(\boldsymbol{y})\left({S_{klpq}^{(\alpha)}(\boldsymbol{y})}-{I_{klpq}^{(\alpha)}}\right)\,d\Theta\right]\mu_{ij}^{(\alpha)}(\boldsymbol{x})\label{eq:34}
\end{equation}
}where $S_{klpq}^{(\alpha)}(\boldsymbol{y})=\int_{\Theta}S_{klmn}(\boldsymbol{y},\tilde{\boldsymbol{y}})N^{(\alpha)}(\tilde{\boldsymbol{y}})\,d\Theta$
and $I_{klmn}^{(\alpha)}(\boldsymbol{y})=I_{klmn}N^{(\alpha)}(\boldsymbol{y})$.
Finally we can express this as a compact form

\begin{equation}
\bar{\boldsymbol{\sigma}}^{\text{}}(\boldsymbol{x})=\mathbf{\bar{L}}:\bar{\boldsymbol{\epsilon}}^{\text{}}(\boldsymbol{x})+\sum_{\alpha=1}^{M}{\bar{\mathbf{M}}}^{(\alpha)}:\boldsymbol{\mu}^{(\alpha)}(\boldsymbol{x})\label{eq:35}
\end{equation}
where $\mathbf{\bar{L}}$ and ${\bar{\mathbf{M}}}^{(\alpha)}$ are
expressed as

\begin{eqnarray}
\mathbf{\bar{L}} & = & \frac{1}{\mid\Theta\mid}\int_{\Theta}\mathbf{L}(\boldsymbol{y}):{\mathbf{E}}(\boldsymbol{y})\,d\Theta\label{eq:36}\\
{\bar{\mathbf{M}}}^{(\alpha)} & = & \frac{1}{\mid\Theta\mid}\int_{\Theta}\mathbf{L}(\boldsymbol{y}):\left({\mathbf{S}^{(\alpha)}(\boldsymbol{y})}-{\mathbf{I}^{(\alpha)}}\right)\,d\Theta\label{eq:37}
\end{eqnarray}

\subsection{Solution to microscale problem}

Eq.(\ref{eq:38}) shows the nonlinear system of equations, written
in incremental form as following

\begin{equation}
\boldsymbol{\Psi}^{(\beta)}={\boldsymbol{\dot{\epsilon}}^{(\beta)}}(\boldsymbol{x})-\sum_{\alpha=1}^{M}\mathbf{S}^{(\alpha\beta)}:\boldsymbol{\dot{\mu}}^{(\alpha)}(\boldsymbol{x})-\mathbf{E}^{(\beta)}:\boldsymbol{\dot{\bar{\epsilon}}}(\boldsymbol{x})=0\label{eq:41}
\end{equation}
and can be solved using Newton's method. Taking derivative of $\boldsymbol{\Psi}^{(\beta)}$
with respect to a variable $\boldsymbol{d}^{(\gamma)}=\boldsymbol{\dot{\epsilon}}^{(\gamma)}$
as

\begin{equation}
\frac{\partial\boldsymbol{\Psi}^{(\beta)}}{\partial\boldsymbol{d}^{(\gamma)}}=\delta_{\beta\gamma}\mathbf{I}-\mathbf{S}^{(\beta\gamma)}:\left(\frac{\partial{\boldsymbol{\dot{\mu}}^{(\beta)}}}{\partial{\boldsymbol{\dot{\epsilon}}^{(\beta)}}}\right)\label{eq:42}
\end{equation}
and further $\frac{\partial{\boldsymbol{\dot{\mu}}^{(\beta)}}}{\partial{\boldsymbol{\dot{\epsilon}}^{(\beta)}}}$
can be calculated as

\begin{equation}
\frac{\partial{\boldsymbol{\dot{\mu}}^{(\beta)}}}{\partial{\boldsymbol{\dot{\epsilon}}^{(\beta)}}}=\mathbf{I}-{{\mathbf{L}}^{(\beta)}}^{-1}:\left(\frac{\partial{\boldsymbol{\dot{\sigma}}^{(\beta)}}}{\partial{\boldsymbol{\dot{\epsilon}}^{(\beta)}}}\right)\label{eq:43}
\end{equation}

Further a coupled damage plasticity constitutive model is used to
find the consistent tangent stiffness, $\frac{\partial{\boldsymbol{\dot{\sigma}}^{(\beta)}}}{\partial{\boldsymbol{\dot{\epsilon}}^{(\beta)}}}$.
For damage model, an actual material with stress state of stress $\boldsymbol{\sigma}$and
strain $\boldsymbol{\epsilon}$ and its fictitious undamaged configuration
with state of stress $\boldsymbol{\widetilde{\sigma}}$ and strain
$\boldsymbol{\widetilde{\epsilon}}$ is considered. This fictitious
state represents an undamaged representative unit with an effective
stress $\boldsymbol{\widetilde{\sigma}}$. The principal of strain
equivalence which ensures equal strains in actual and fictitious configurations
i.e. $\boldsymbol{\epsilon}=\boldsymbol{\widetilde{\epsilon}}$ gives
the relationship between actual and effective stress as

\begin{equation}
{\boldsymbol{\sigma}}^{(\beta)}={\boldsymbol{\widetilde{\sigma}}}^{(\beta)}(1-\omega^{(\beta)})\label{eq:44}
\end{equation}
where $\omega$ is a scalar damage variable and defines the ratio
of damaged to actual area of representative unit of the material.
Plasticity is associated with the undamaged portion of representative
unit which leads the following constitutive relation between effective
stress and elastic strain:

\begin{equation}
{\boldsymbol{\widetilde{\sigma}}}^{(\beta)}={{\mathbf{L}^{e}}^{(\beta)}}:{{\boldsymbol{\epsilon}}^{e}}^{(\beta)}\label{eq:45}
\end{equation}

Eq. (\ref{eq:44}) and (\ref{eq:45}) can be combined as:

\begin{equation}
{\boldsymbol{\sigma}}^{(\beta)}=(1-\omega^{(\beta)}){{\mathbf{L}^{e}}^{(\beta)}}:{{\boldsymbol{\epsilon}}^{e}}^{(\beta)}\label{eq:44-1}
\end{equation}
and its rate form as

\begin{equation}
{\boldsymbol{\dot{\sigma}}}^{(\beta)}=(1-\omega^{(\beta)}){{\mathbf{L}^{e}}^{(\beta)}}:{{\boldsymbol{\dot{\epsilon}}}^{e}}^{(\beta)}-\dot{\omega}^{(\beta)}{\boldsymbol{\widetilde{\sigma}}}^{(\beta)}\label{eq:44-1-1}
\end{equation}

A damage evolution function, $f_{D}$ is defined in strain space for
damage evolution as 

\begin{equation}
f_{D}^{(\beta)}=\epsilon_{D}^{(\beta)}-\kappa_{D}^{(\beta)}\label{eq:48}
\end{equation}
For defining the damage equivalent strain, $\epsilon_{D}$ a maximum
principal strain based criteria can be used 

\begin{equation}
\epsilon_{D}^{(\beta)}=\max\left\{ \langle\hat{\boldsymbol{\epsilon}}\rangle^{(\beta)}\right\} =\max\left\{ \langle\hat{{\epsilon}}_{1}\rangle^{(\beta)},\langle\hat{{\epsilon}}_{2}\rangle^{(\beta)},\langle\hat{{\epsilon}}_{3}\rangle^{(\beta)}\right\} \label{eq:52}
\end{equation}
where $\hat{\boldsymbol{\epsilon}}$ is the principal values of ${\boldsymbol{\epsilon}}^{(\beta)}$.
Damage growth conditions are as follows:

\begin{equation}
\dot{\kappa}_{D}^{(\beta)}\geq0,\qquad f_{D}^{(\beta)}\leq0,\qquad\dot{\kappa}_{D}^{(\beta)}f_{D}^{(\beta)}=0\label{eq:53-1}
\end{equation}

Damage growth increment is related with total strain as follows:

\begin{equation}
d\omega^{(\beta)}=\left(\frac{d\omega}{d\kappa_{D}}\right)^{(\beta)}\left(\frac{d\epsilon_{D}}{d\boldsymbol{\epsilon}}\right)^{(\beta)}d\boldsymbol{\epsilon}^{(\beta)}\label{eq:53-2}
\end{equation}

The scalar damage parameter, $\omega^{(\beta)}$ can be expressed
in terms of a deformation history parameter, $\kappa_{D}^{(\beta)}$
as an exponential function:

\begin{equation}
\omega^{(\beta)}=1-exp\left(\frac{1}{m^{(\beta)}}\left(1-\left(\frac{{\kappa_{D}^{(\beta)}}}{\kappa_{Df}^{(\beta)}}\right)^{m^{(\beta)}}\right)\right)\label{eq:55}
\end{equation}
$\kappa_{Df}^{(\beta)}$ corresponds to deformation history parameter
at state of full damage. $m^{(\beta)}$ is the material constant which
can be calculated using \textcolor{black}{the experimental stress-strain
data of each phase}. \textcolor{black}{A large value of $m$ signifies
the brittle nature of the material where as small value of $m$ represents
the ductile behavior of the material.}

\begin{equation}
\dot{\omega}^{(\beta)}=\frac{1}{2\kappa_{Df}^{(\beta)}}\left(1-\omega^{(\beta)}\right)\left(\frac{{\kappa_{D}^{(\beta)}}}{\kappa_{Df}^{(\beta)}}\right)^{m^{(\beta)}-1}\dot{\kappa}_{D}^{(\beta)}\label{eq:55-1}
\end{equation}

Similarly plastic potential function is defined in effective stress
space as 

\begin{equation}
f_{p}^{(\beta)}=\sigma_{eq}^{(\beta)}({\boldsymbol{\widetilde{\sigma}}})-\sigma_{Y}^{(\beta)}(\kappa_{p})\label{eq:48-1}
\end{equation}
where $\sigma_{eq}$ is equivalent stress derived from classical yield
functions (e.g. von-Mises or Rankine) in effective stress space. $\sigma_{Y}$
is yield stress which depends on equivalent plastic strain, $\kappa_{P}$.
This yield criteria also satisfy Kuhn-Tucker conditions for loading/unloading:

\begin{equation}
\dot{\lambda}^{(\beta)}\geq0,\qquad f_{p}^{(\beta)}\leq0,\qquad\dot{\lambda}_{p}^{(\beta)}f_{p}^{(\beta)}=0\label{eq:53-1-1}
\end{equation}
where $\lambda$ is a plastic multiplier used for classical flow rule.
Total strain can be decomposed into elastic and plastic strain components
by using standard additive decomposition assumption which gives the
elastic strain rate as: 

\begin{equation}
{{\boldsymbol{\dot{\epsilon}}}^{e}}^{(\beta)}={\boldsymbol{\dot{\epsilon}}}^{(\beta)}-\dot{\lambda}^{(\beta)}{\boldsymbol{n}^{(\beta)}(\boldsymbol{\widetilde{\sigma}})}\label{eq:55-2}
\end{equation}
$\boldsymbol{n}$ is a vector normal to yield surface defined in effective
stress space. Finally the tangent stiffness for the phase partition
(as defined by Eq.(\ref{eq:44-1-1})) is updated by using Eq.(\ref{eq:53-2}),
(\ref{eq:55-1}) and (\ref{eq:55-2}).

\subsection{Determination of elastic and phase damage coefficient tensors\label{subsec:Influence-functions}}

Using Eq.(\ref{eq:30}), Eq.(\ref{eq:25}) can be simplified by writing
it for particular sub-domain

\begin{equation}
\left\{ L_{ijkl}(\boldsymbol{y})\left(E_{klmn}(\boldsymbol{y})\bar{\epsilon}_{mn}(\boldsymbol{x})+\sum_{\alpha=1}^{M}\left(S_{klpq}^{(\alpha)}(\boldsymbol{y})-I_{klpq}^{(\alpha)}(\boldsymbol{y})\right)\mu_{kl}^{(\alpha)}(\boldsymbol{x})\right)\right\} _{,y_{j}}=0\label{eq:58}
\end{equation}

Eq.(\ref{eq:58}) is valid for any arbitrary macroscopic fields such
as
\begin{itemize}
\item $\bar{\epsilon}_{mn}(\boldsymbol{x})\neq0$ where as $\mu_{mn}^{(\alpha)}(\boldsymbol{x})=0$,
which leads to
\begin{equation}
\left\{ L_{ijkl}(\boldsymbol{y})E_{klmn}(\boldsymbol{y})\right\} _{,y_{j}}=0\label{eq:59}
\end{equation}
\item $\bar{\epsilon}_{mn}(\boldsymbol{x})=0$ where as $\mu_{mn}^{(\alpha)}(\boldsymbol{x})\neq0$,
which leads to
\end{itemize}
\begin{equation}
\left\{ L_{ijkl}(\boldsymbol{y})\left[\sum_{\alpha=1}^{M}\left(S_{klpq}^{(\alpha)}(\boldsymbol{y})-I_{klpq}^{(\alpha)}(\boldsymbol{y})\right)\right]\right\} _{,y_{j}}=0\label{eq:60}
\end{equation}

Theses two cases represent two distinct boundary value problems which
can be used to calculate $\mathbf{E}$ and $\mathbf{S}$. Elastic
coefficient tensor, $\mathbf{E}$ can be determined by solving a FE
problem for a RVE as defined by Eq.(\ref{eq:59}) along with applied
periodic boundary conditions. The solution of this leads to Eq.(\ref{eq:38}),
which shows that the elastic coefficient tensor for an element is
the measure of strain in each element due to applied unit strain at
macrolevel in the absence of any eigen strain. ${E}_{ijkl}$ is a
fourth order tensor which can be suitably converted to $6\times6$
matrix for 3-dimensional stress state. So applied strain in each direction
corresponds to the column of elastic coefficient matrix for each element.
\textcolor{black}{In the present manuscript, all the verification calculations
are performed using 2-dimensional plane strain based FE approach.
Extension to 3-dimensional approach is merely straightforward. For
2-D plane strain element, the following expression depicts the case
for the unit normal strain applied in 1-direction:}

\textcolor{black}{
\begin{equation}
\left[\begin{array}{c}
\epsilon_{11}\\
\epsilon_{22}\\
\epsilon_{12}
\end{array}\right]=\left[\begin{array}{ccc}
E_{11} & E_{12} & E_{16}\\
E_{21} & E_{22} & E_{26}\\
E_{61} & E_{62} & E_{66}
\end{array}\right]\times\left[\begin{array}{c}
1\\
0\\
0
\end{array}\right]\label{eq:61}
\end{equation}
}

Similarly by applying strains one by one in other directions, the
total $\mathbf{E}$ matrix can be evaluated for each element. After
obtaining the elastic coefficient tensor, $\mathbf{E}$, the effective
moduli $\bar{\mathbf{L}}$ can be calculated as per Eq.(\ref{eq:37}).
Similarly phase damage coefficient tensor, $\mathbf{S}$ can be determined
by solving a FE problem for a RVE as defined by Eq.(\ref{eq:60})
along with applied periodic boundary conditions. The phase damage
coefficient tensor can be defined as the measure of strain induced
in an element due to unit strain applied in another element in the
domain. Similarly this tensor can also be written as $6\times6$ matrix.
It would not be possible to apply the unit strain on an irregular
element so the nodal forces corresponding to unit strain are calculated
and applied on each element one by one. The measured strains in the
other elements gives the phase damage influence function as expressed
below for applied unit strain in 1-direction. \textcolor{black}{This
can be expressed as following for 2-D plane strain element:}

\textcolor{black}{
\begin{equation}
\left[\begin{array}{c}
\epsilon_{11}\\
\epsilon_{22}\\
\epsilon_{12}
\end{array}\right]=\left[\begin{array}{ccc}
S_{11} & S_{12} & S_{16}\\
S_{21} & S_{22} & S_{26}\\
S_{61} & S_{62} & S_{66}
\end{array}\right]\times\left[\begin{array}{c}
1\\
0\\
0
\end{array}\right]\label{eq:62}
\end{equation}
}

Similarly the matrix can be fully populated by evaluating the strain
for the other cases.

\section{Computational Aspects\label{sec:Computational-Aspects}}

\subsection{Strain localization\label{subsec:Strain-localization}}

Discretization of microscale domain into number of partitions can
induce the localization of inelastic process into an arbitrary zone
of particular dimension. This leads to pathological sensitivity of
numerical results to the size of partition. \textcolor{black}{This problem can be efficiently solved by using gradient based regularization technique i.e. by using stain gradient or damage gradient where an additional information around a point's neighborhood is provided for the damage description \cite{placidi2018strain, placidi2016variational}.} Other remedy of this problem is nonlocal regularization technique, which is also explained earlier in section 2.1, using a specific form of distribution function,
$\varphi$. By selecting a particular shape of this function along
with size of interaction radius the effect of adjacent partitions
can be accounted for in the calculation of strain in a particular
partition. Use of a complex function increases the computation time
and efforts which makes this solution less attractive. Therefore a
simple form of distribution function:

\begin{equation}
\varphi(\boldsymbol{y}-\mathbf{\zeta})=\begin{cases}
1 & \boldsymbol{y}\in\Theta^{(\beta)}\\
0 & \boldsymbol{y}\notin\Theta^{(\beta)}
\end{cases}\label{eq:63}
\end{equation}
and

\begin{equation}
\bar{\varphi}^{(\beta)}=\frac{\varphi(\boldsymbol{y}-\mathbf{\zeta})}{\int\varphi(\boldsymbol{y}-\mathbf{\zeta})\,d\zeta}=\frac{1}{\mid\Theta^{(\beta)}\mid}\label{eq:64}
\end{equation}
is often chosen. Using this form of nonlocal distribution function
may fail to alleviate the problem of strain localization. 

\begin{figure}[H]
\begin{centering}
\includegraphics[width=0.75\textwidth]{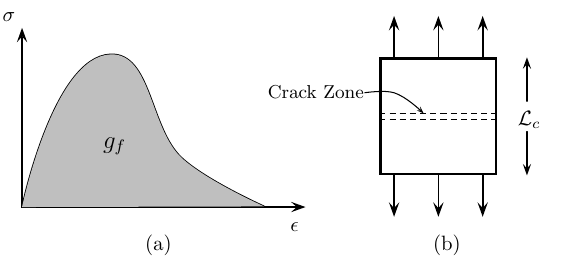}
\par\end{centering}
\caption{(a). Energy dissipated per unit volume, '$g_{f}$' is calculated as
area under stress strain diagram (b). characteristic length '$\mathcal{L}_{c}$'
is the height of a rectangular region aligned in the direction of
applied load. \label{fig:4-1}}
\end{figure}

Another remedy to this problem is based on the appropriate adjustment
of certain model parameters that control the softening depending upon
the size of partition. This method is proposed and validated by adjusting
these model parameters and results are discussed in section \ref{sec:Numerical-implementation-procedu}.
In this approach, a characteristic dimension, '$\mathcal{L}_{c}$'
is used which is also called as width of the crack band or the damage
localized region in continuum damage model. For each constituent,
the area under the stress strain diagram,

\begin{equation}
g_{f}=\int_{0}^{\infty}\sigma(\epsilon)\,d\epsilon\label{eq:65-1}
\end{equation}
represents the energy dissipated per unit volume at complete failure
(see Fig. \ref{fig:4-1}). 

The fracture energy of the material, $\mathcal{G}_{F}$ which is generally
determined from mode-I fracture experiments is related to $\mathcal{L}_{c}$
as

\begin{equation}
\mathcal{G}_{F}=\mathcal{L}_{c}g_{f}\label{eq:66-1}
\end{equation}

In case of a rectangular region of isotropic material, $\mathcal{L}_{c}$
has a direction aligned with applied load and equal to the height
of that region as shown in Fig. \ref{fig:4-1}. So according to the
size of partition the stress strain relation is corrected to make
the fracture energy independent of the size of partition. This simple
looking remedy has some limitations that the crack growth path should
be parallel to one of the edge and shape of the region should be structured.
\citet{oliver1989consistent} proposed a method to overcome this problem
for shell element subjected to the multiaxial loading conditions.
\citet{oliver1989consistent} showed that the characteristic length
can be estimated as the reciprocal of partial derivative of an auxiliary
function, '$\Phi$' with respect to a direction normal to crack band
as:

\begin{equation}
\mathcal{L}_{c}=\left(\frac{\partial\Phi}{\partial\boldsymbol{x}'}\right)^{-1}=\left(\sum_{i=1}^{n_{c}}\Phi_{i}\frac{\partial N_{i}}{\partial\boldsymbol{x}'}\right)\label{eq:67-2}
\end{equation}

In a transformed coordinate system $\boldsymbol{x}'$ represents a
direction vector normal to crack band. $n_{c}$ is the number of corners,
$N_{i}$ are the shape functions and $\Phi_{i}$ is the value of $\Phi$
at corner '$i$'. The values of $\Phi_{i}$ are determined by passing
a line through the centroid of the partition and aligned along the
possible crack band direction. For the corners on one side of this
crack line the values of $\Phi_{i}$ are set to '$0$' and on the
other side as '$1$'. There are various methods suggested in literature
\citep{pineda2012implementation,vcervenka2005equivalent,oliver1989consistent}
to determine the direction \textcolor{black}{of} expected crack band. One method proposed
by \citet{oliver1989consistent} is based on localization analysis
and uses normal to direction of vanishing determinant of acoustic
tensor, $\boldsymbol{n.\mathbf{D}_{ed}.n}$ where $\boldsymbol{\mathbf{D}_{ed}}$
is the stiffness tensor of the damaged model. 

\begin{figure}[H]
\begin{centering}
\includegraphics[width=0.85\textwidth]{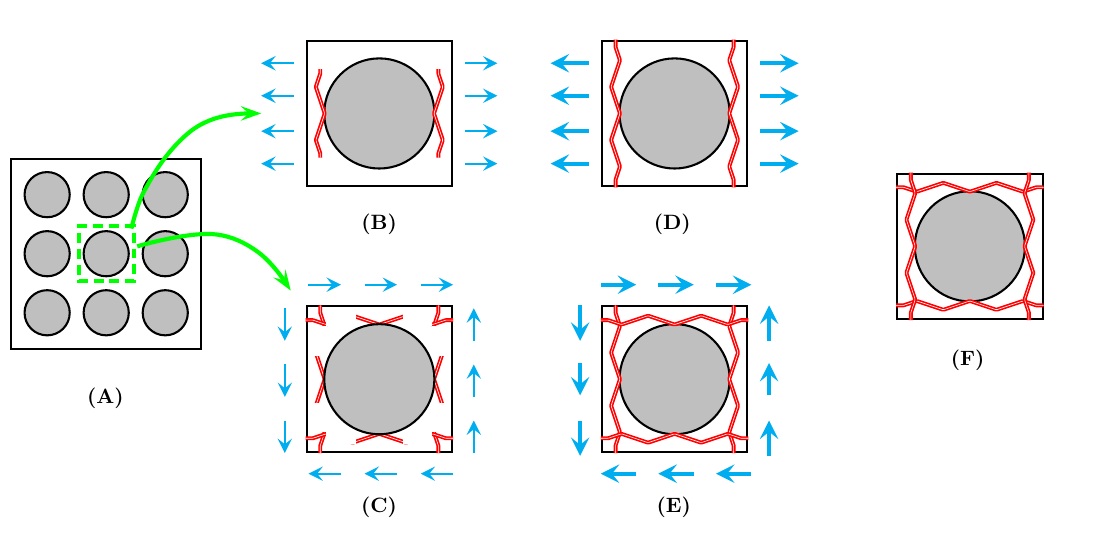}
\par\end{centering}
\centering{}\caption{Failure path map for RVE subjected to multiaxial loading conditions.
A RVE is identified from a fiber-matrix network (A). As soon as the
tensile (B) or shear (C) load reaches at critical level, crack starts
to form in the matrix region. The crack path map is obtained by further
increasing the tensile/compressive (D) and shear (E) loads till the
complete matrix failure occur. Finally all the crack paths are superimposed
(F) representing all the possible failure paths under the state of
multiaxial load.\label{fig:5}}
\end{figure}

Other method is by aligning the crack band along the normal to maximum
principal strain/stress which also matches with the failure criteria
used in the present formulation (see Eq. (\ref{eq:52})). The direction
of maximum principal stress/strain for a region under uniaxial loading
conditions can be determined by various methods. Cracking in each
loading case represents the dominant failure mode when subjected to
multiaxial loads. For a domain with an inclusion it would be difficult
to determine the direction of principal strain which can be found
by using FE methods. In the present method, crack band direction is
determined by performing uniaxial FE analyses on a RVE and failure
paths are drawn corresponding to each load case. Later all the failure
paths are superimposed to get the failure path map of RVE subjected
to multiaxial loading. Fig. \ref{fig:5} shows the overall scheme
of obtaining the crack band direction where (A) refers to the identification
of RVE out of periodic array of fibers which embeds in a matrix. This
RVE is tested under (B) tensile and (C) shear loads and failure starts
as soon these loads reach to critical level. Further increase of loading
leads to complete matrix failure and crack paths are found as shown
in Fig. \ref{fig:5}(D) and (E). Finally all the results are superimposed
to get a map of all the failure paths.

After getting the value of $\mathcal{L}_{c}$, the value of $g_{f}$
is calculated from the area under the stress-strain diagram. As per
Eq. (\ref{eq:55}), the softening curve of stress can be represented
as function of material constant, $m$. By varying the value of $m$
the area under the stress-strain diagram can be adjusted to get the
desired value of $g_{f}$ so that the fracture energy $\mathcal{G}_{F}$
becomes invariant with respect to the size of partition. 

\subsection{Partitioning Strategy }

It has been discussed that the model reduction strategy plays an important
role in the accurate prediction of macroscale results. The coarse
representation of eigen strains gives rise to inaccurate macroscale
results . Computationally economical way of one partition per phase
results in an artificial post failure stiffness \citep{singh2017reduced}
which can be alleviated by increasing the number of partitions. One
partition per element is always proven as computationally expensive
solution. Other ways of alleviating this shortcoming such as use of
higher order shape functions \citep{michel2003nonuniform}, dynamic
partitioning scheme \citep{oskay2007eigendeformation}, hybrid compatible-incompatible
eigen strain field \citep{fish2013hybrid} are also available in literature
and can be used at the cost of increase in the complexity. Otherwise,
in order to get accurate macroscale predictions, decision regarding
the optimum number of partitions is always seen as a challenge. This
issue can be solved by performing the comparative studies for a RVE
with different number of partitions. Apart from the number of partitions,
the selection regarding partitioning area is also a key. Merely increasing
the number of partitions does not guarantee an improvement of results.
\textcolor{black}{Identification of optimal reduced order model in
terms of partitioned sub-domain selection has been studied by \citet{sparks2013identification}
where an optimization technique was proposed to minimize the modeling
related errors. This method uses a genetic algorithm for optimization
by minimizing the difference of two response metrics i.e. one related
with full resolution of the microstructure and other with reduced
order model. Although this optimization procedure was proved effective
in the identification of satisfactory partitions, still computational
exhaustiveness in case of densely meshed microstructure for searching
of an optimal space could not be ignored which makes this as an element
size dependent formulation. }

\textcolor{black}{Other partitioning method based on the failure modes
of the micro-constituents was suggested by \citet{bogdanor2017prediction}.
Various damage mechanisms i.e. fiber fracture (tensile), fiber crushing
(compressive), matrix cracking or crushing and delaminations were
pre-identified and sub-domains for each phase were created to capture
each damage mode accordingly.}

\begin{figure}[H]
\centering{}\includegraphics[width=0.6\textwidth]{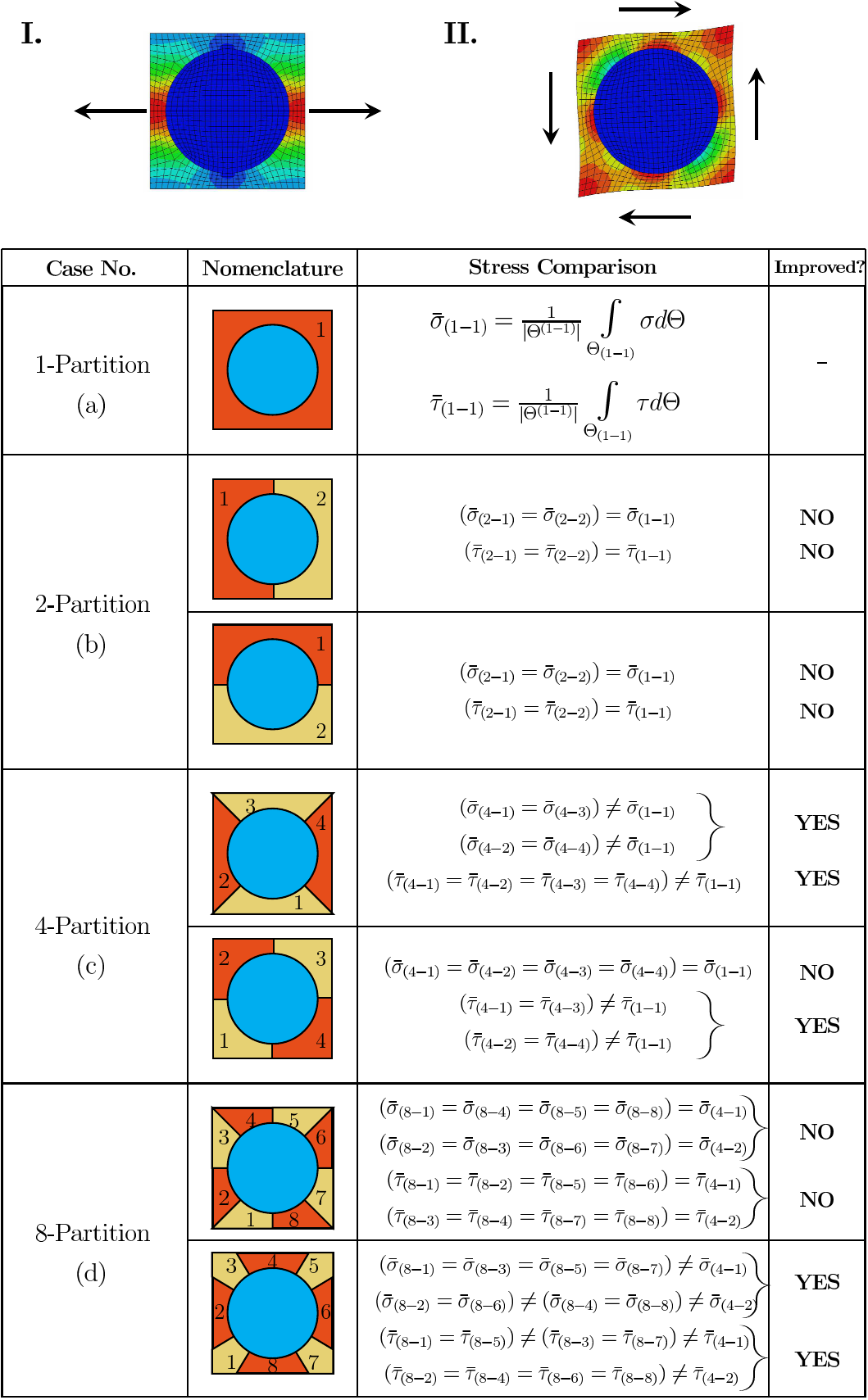}\caption{Partitioning the matrix domain based on eigen strain/stress distribution
of RVE subjected to tension and shear load. Single partition per phase
average partition stress is compared with two and four partition results.
Similarly four and eight partition results are further used for deciding
the final shape of eight partitions. Nomenclature used is $\bullet_{(x-y)};$
$x=$Total number of partitions and $y=$Particular partition number.\label{fig:6} }
\end{figure}
A novel strategy regarding the partitioning of RVE domain is presented
in this manuscript. This strategy is based on the the nonuniform distribution
of eigen strain in a RVE when it is subjected to multi axial loading
conditions and more applicable to those solutions where $C^{0}$ shape
functions are used to estimate the eigen strain in a partition or
uniform distribution of eigen strain is assumed. For illustration,
this procedure is shown in Fig. \ref{fig:6} for a single fiber RVE
and the stress distribution is obtained from uniaxial and pure shear
loading conditions. 

First of all computationally economical single partition averaged stress
or strain data is calculated. Next we can compare this averaged data
of a partition for single partition per phase model with another model
where two partitions for matrix are being utilized. If the average
stress per partition remains same for both the tensile and shear load
cases then it is required to change the number or shape of partitions
for the matrix. This has been validated for three different cases
where two (2-Partition), four (4-Partition) and eight (8-Partition)
matrix partitions are considered for comparison with single partition
average. Two different ways of partitioning are adopted for each case
which results either in different shape or position of partitioning
zone as shown in Fig. \ref{fig:6}. No improvement is observed for
2-Partition model (see Fig. \ref{fig:6}(b)) since the tensile and
shear stress/strain average for each partition remains same as 1-Partition
model (see Fig. \ref{fig:6}(a)). Similarly those 4-Partition and
8-Partition models are selected which lead to different tensile and
shear stress/strain average for each partition than 1-Partition and
4-Partition model respectively. These selected models could be used
further for the macroscale predictions.  Same arguments are valid
if we discretize the fiber domain also. 

\section{Numerical implementation procedure\label{sec:Numerical-implementation-procedu}}

\begin{figure}[H]
\centering{}\includegraphics[width=0.5\textwidth]{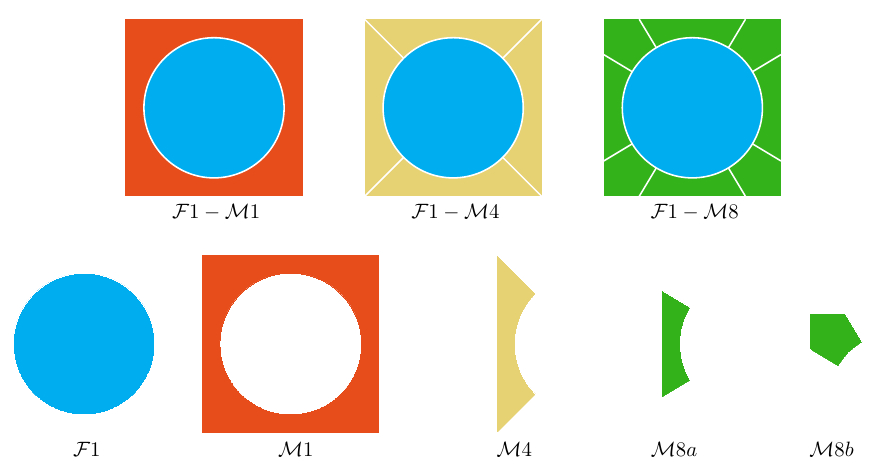}\caption{Nomenclature for three different partition models; In all three models,
single partition is used for fiber zone and named as $\mathcal{F}1$
and matrix partition is named as $\mathcal{M}1$ for first model.
For second model, matrix is partitioned into four zone and each one
is named as $\mathcal{M}4$. For third model, matrix is divided into
eight zones and results two different types of partitions which are
named as $\mathcal{M}8$a and $\mathcal{M}8$b.\label{fig:9}}
\end{figure}

The overall numerical implementation procedure can be divided into
two stages. First stage is named as ``preprocessing stage'' which
includes following substeps:
\begin{enumerate}
\item Creation of RVE.
\item Partitioning of RVE domain for order reduction.
\item Calculation of coefficient tensors.
\end{enumerate}
Coefficient tensors can be calculated either by using finite element
method as explained in section \ref{subsec:Influence-functions} or
by analytical method \citep{dvorak2001transformation}. Preprocessing
stage gives $\bar{\mathbf{L}}$ and $\bar{\mathbf{M}}^{(\alpha)}$
which can be used further to solve macroscale problem. The second
stage is about the evaluation of problem defined at macroscale and
named as ``solution stage''. This problem is also generally solved
by using finite element method. This stage consists of following substeps:
\begin{enumerate}
\item Determination of characteristic length for each partition.
\item Modification of stress-strain relationship for each partition using
material parameter '$m$' based on calculated characteristic length.
\item Calculation of eigen strain at each integration point using corresponding
macroscopic strain. This eigen strain is contributed by state of damage
and plastic strain corresponding to that integration point.
\item Calculation of macroscale stress using macroscale strain, eigen strain
and coefficient tensors.
\end{enumerate}
By using above mentioned steps, this section demonstrates the numerical
implementation procedure by validating the proposed formulation for
a RVE of unit length with a single circular fiber embedded at the
center (as shown Fig. \ref{fig:6}). The macroscale problem using
the calculated parameters is solved as explained in section \ref{sec:Macoscale-Simulation}.
Material data corresponding to E-glass/Epoxy is used for the constituents.
The RVE simulations for influence functions are performed using a
commercial software ABAQUS. For easy implementation of periodic boundary
conditions, it is necessary to match each master node exactly with
the slave node on the corresponding faces which requires a structured
mesh for the RVEs. The constraint equations defined as 

\begin{equation}
\boldsymbol{u}\mid_{(\boldsymbol{y}=0)}=\boldsymbol{u}\mid_{(\boldsymbol{y}=\boldsymbol{y^{0}})}+\boldsymbol{\epsilon y^{0}}
\end{equation}
are used for periodic boundary conditions where the slave nodes are
tied with master nodes at all the edges and corners of RVE. For checking
the effect of partitions and their associated characteristic dimensions,
RVE studies are performed using three different partition strategies
as shown in Fig. \ref{fig:9}. First one considers single partition
for fiber and matrix domain. Second consists of single partition for
fiber and four partitions for matrix domain where as in third case
eight matrix partitions with single fiber partition are taken. These
cases are given a nomenclature as $\mathcal{F}1-\mathcal{M}1$, $\mathcal{F}1-\mathcal{M}4$
and $\mathcal{F}1-\mathcal{M}8$ respectively. 

\subsection{Calculation of coefficient tensors}

\begin{figure}[H]
\begin{centering}
\begin{minipage}[t]{0.17\textwidth}%
\begin{center}
\includegraphics[bb=0bp 120bp 636bp 535bp,width=0.75\textwidth]{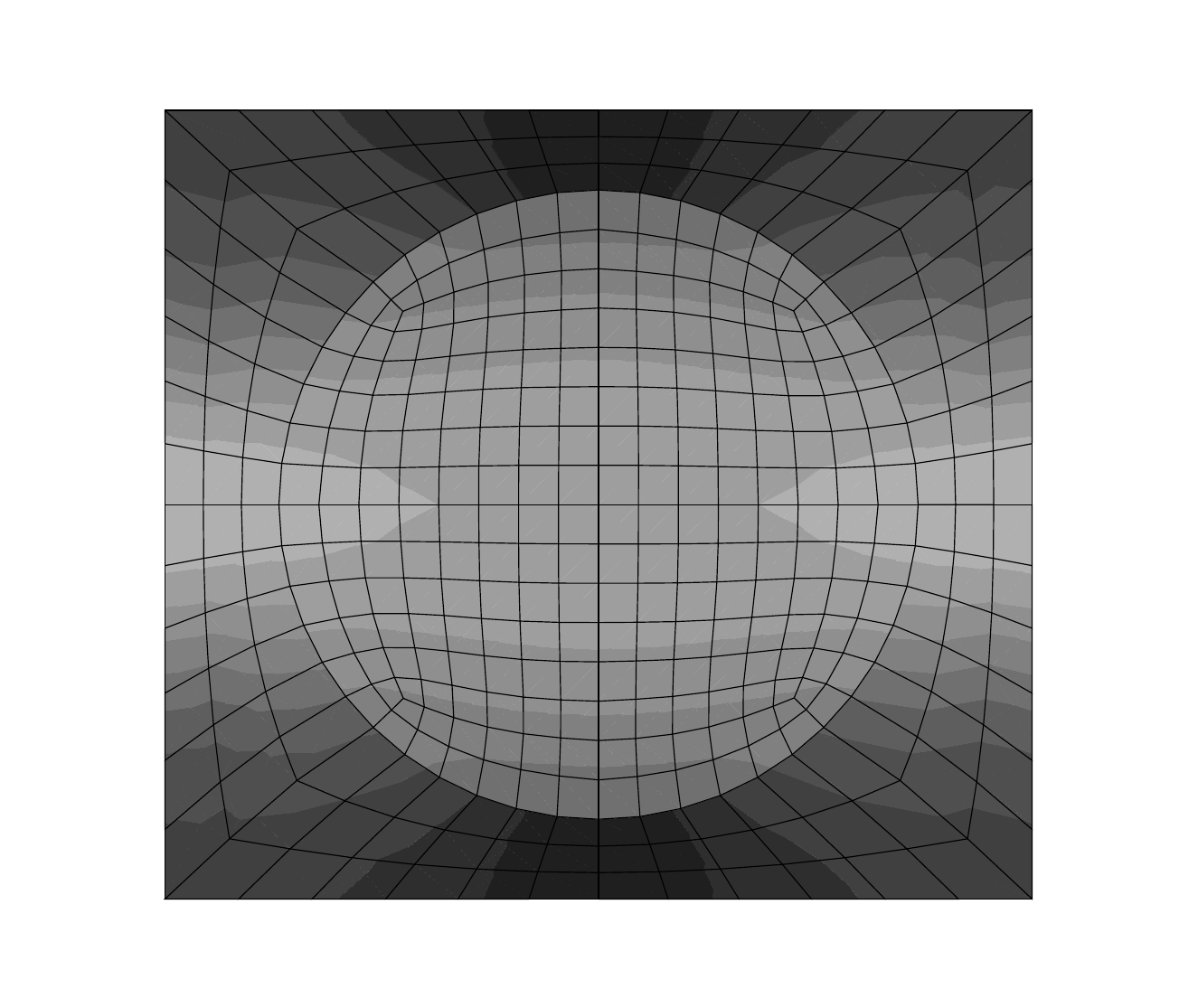}
\par\end{center}
\begin{center}
(a). $\epsilon_{11}$
\par\end{center}%
\end{minipage}\hspace{1.00cm}%
\begin{minipage}[t]{0.17\textwidth}%
\begin{center}
\includegraphics[bb=0bp 120bp 636bp 535bp,width=0.75\textwidth]{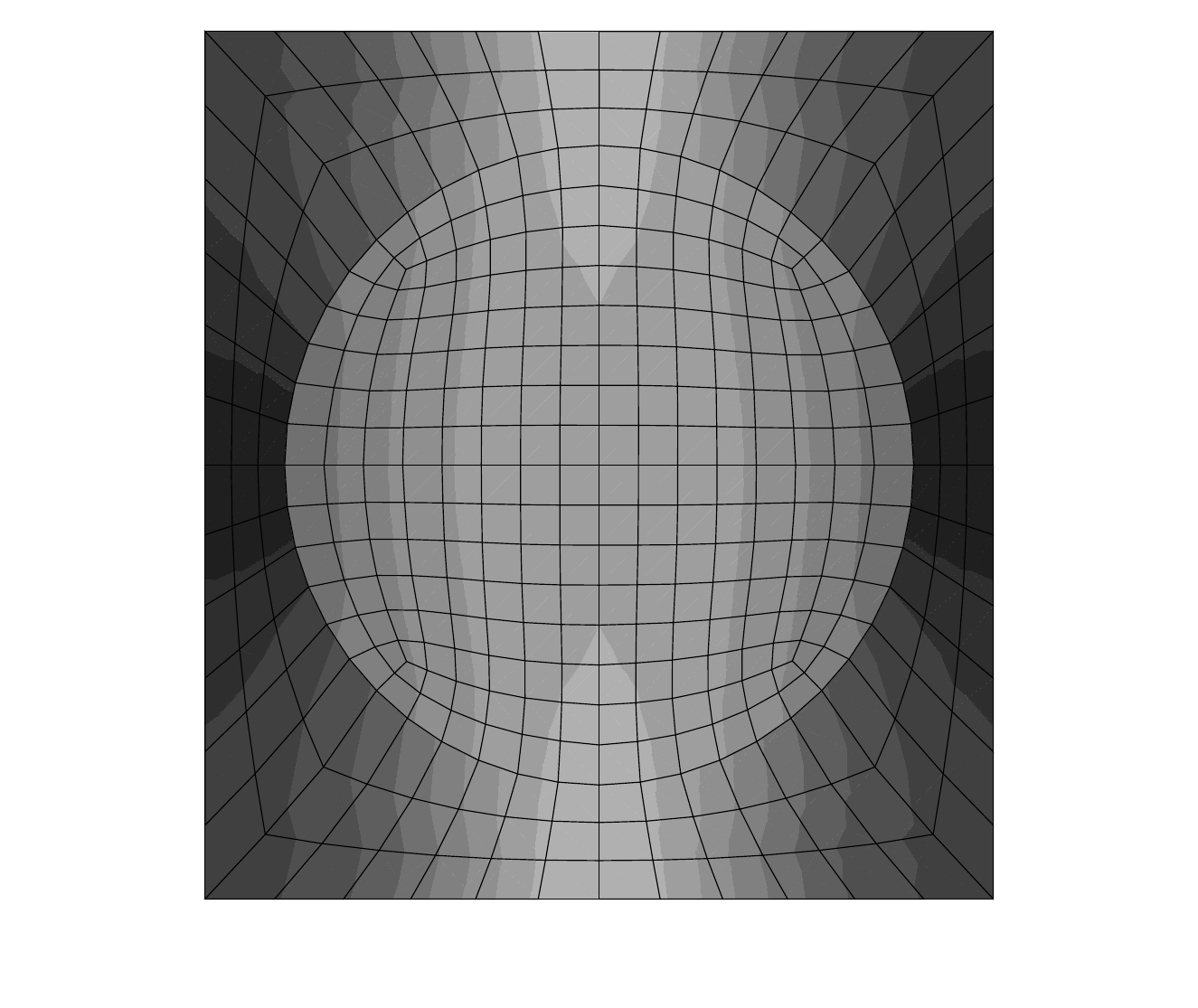}
\par\end{center}
\begin{center}
(b). $\epsilon_{22}$
\par\end{center}%
\end{minipage}\hspace{1.00cm}%
\begin{minipage}[t]{0.17\textwidth}%
\begin{center}
\includegraphics[bb=0bp 120bp 636bp 535bp,width=0.75\textwidth]{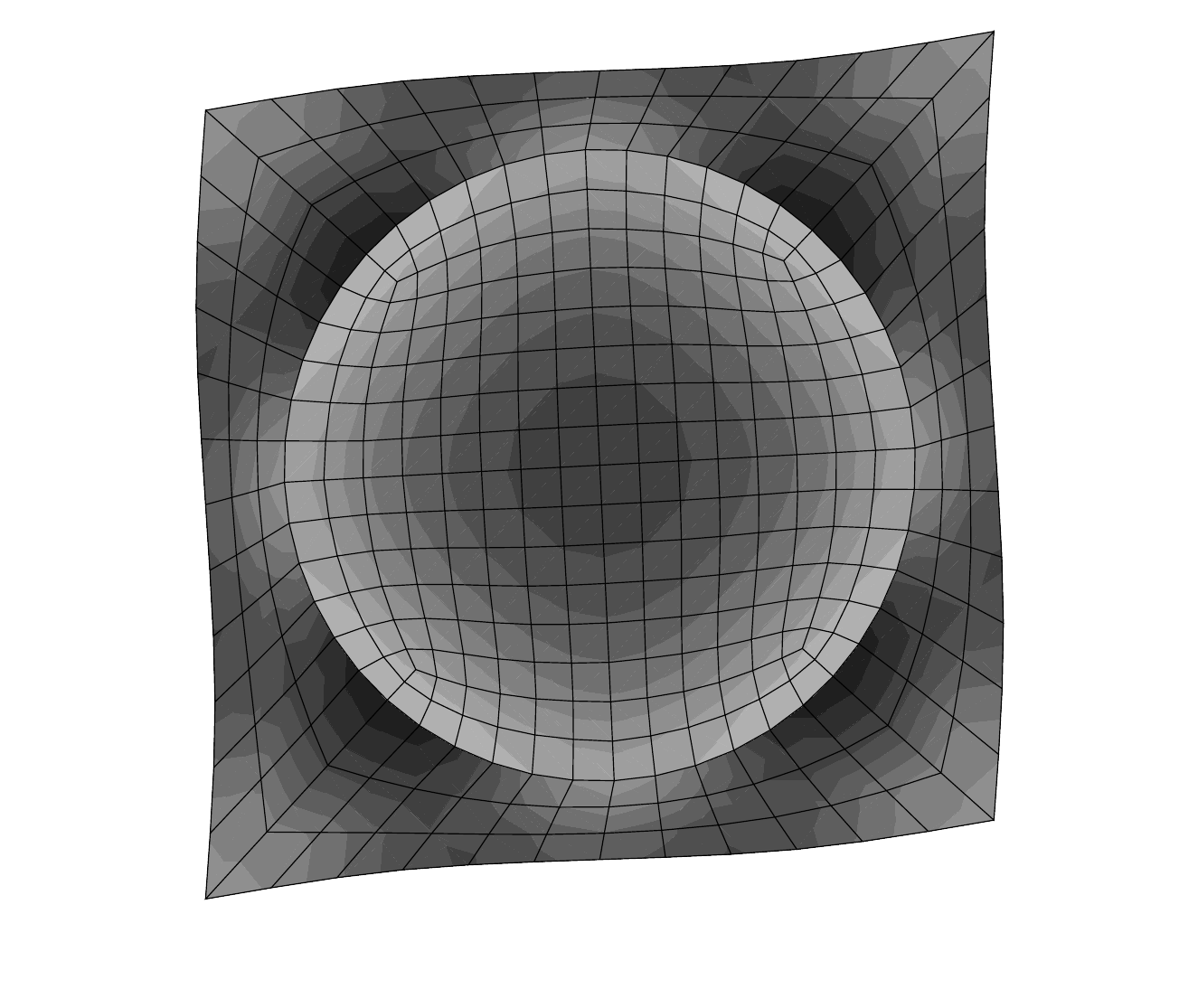}
\par\end{center}
\begin{center}
(c). $\epsilon_{12}$
\par\end{center}%
\end{minipage}
\par\end{centering}
\centering{}\caption{Elastic coefficient tensor represented as strain response under three
distinct applied load vectors (a). longitudinal tension (b). transverse
tension and (c). in-plane shear.\label{fig:7}}
\end{figure}

The elastic influence function represents the relation between the
macroscopic strain to microscopic strain and signifies the amount
of heterogeneity in RVE or material. Fig. \ref{fig:7} shows the directional
strain plots for three load vectors corresponding to each column of
the elastic coefficient matrix. Each response for a particular load
vector also represent the eigen deformation mode shape.

Following a similar procedure, the phase damage coefficient tensor,
which represents the relationship between macroscopic strain to applied
eigen strain is calculated. 

\subsection{Calculation of characteristic lengths}

Based on the shape and size of partition, the characteristic length
as explained in section \ref{subsec:Strain-localization} is calculated
for each case i.e. $\mathcal{M}4$ and $\mathcal{M}8$. $\mathcal{M}4$
consists of partitions of equal size and $\mathcal{M}8$ consists
of eight partitions of two different sizes which are named as $\mathcal{M}8a$
and $\mathcal{M}8b$ as shown in Fig. \ref{fig:9}.

\begin{figure}[H]
\centering{}\includegraphics[width=0.75\textwidth]{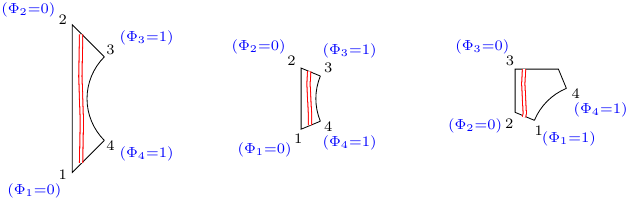}\caption{Crack path marked for different partitions (a). $\mathcal{M}4$ (b).
$\mathcal{M}8a$ and (c). $\mathcal{M}8b$ with auxiliary function
'$\Phi$' values marked for each corner.}
\end{figure}

For each partition the crack path is drawn based on the failure modes
of a unit cell when subjected to various loading conditions as shown
in Fig. \ref{fig:5}. $\mathcal{L}_{c}$ for each three sizes are
calculated using following expression:

\begin{equation}
\mathcal{L}_{c}=\left(\sum_{i=1}^{n_{c}}\left[\frac{\partial N_{i}}{\partial x}\cos\theta+\frac{\partial N_{i}}{\partial y}\sin\theta\right]\Phi_{i}\right)^{-1}\label{eq:67-1}
\end{equation}
where $\theta$ is the angle between $x$-axis and normal to crack
plane. Auxiliary function $\Phi$ is taken as $'1'$ for the corners
in the positive $x$-direction and $'0'$ elsewhere. 

\begin{figure}[H]
\centering{}%
\begin{minipage}[t]{0.4\textwidth}%
\begin{center}
\caption{Variation of stress with respect to strain for different matrix partitons
$\mathcal{M}1$, $\mathcal{M}4$, $\mathcal{M}8$a and $\mathcal{M}8$b
(a). using same characteristic lengths (b). using different characteristic
lengths and without consideration of plasticity and (c). using different
characteristic lengths and with consideration of plasticity.\label{fig:stress-strain}}
\par\end{center}%
\end{minipage}\quad{}%
\begin{minipage}[t]{0.54\textwidth}%
\begin{center}
\includegraphics[width=0.8\textwidth]{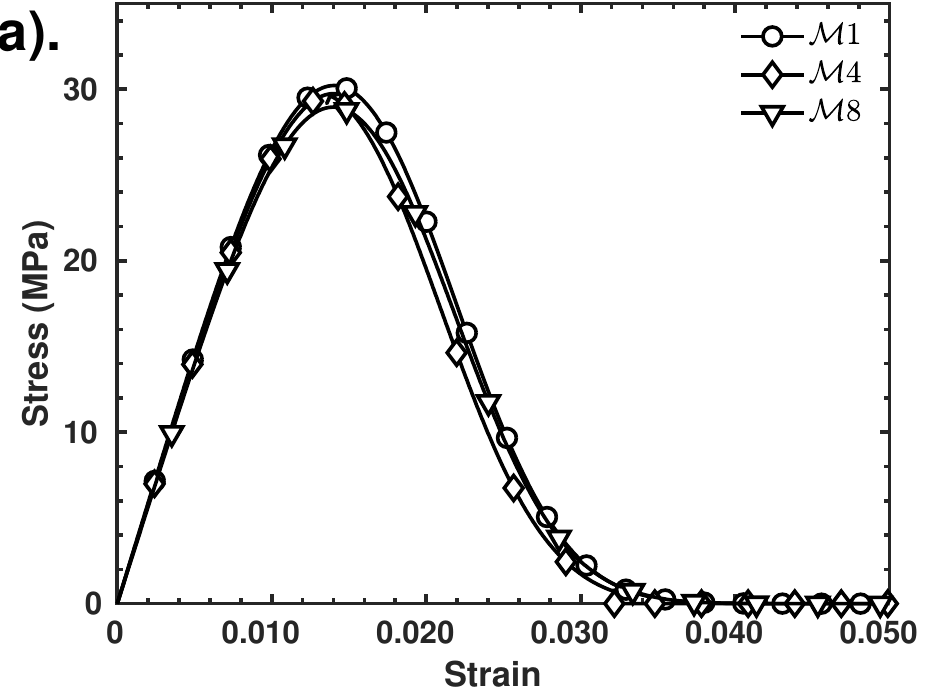}
\par\end{center}
\vspace{0.1cm}
\begin{center}
\includegraphics[width=0.8\textwidth]{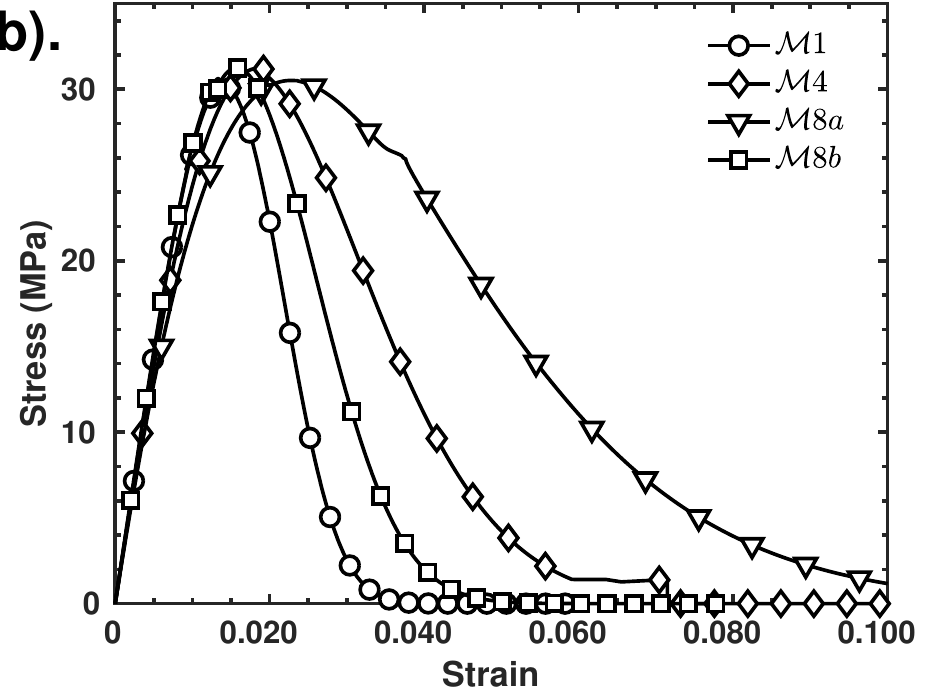}
\par\end{center}
\vspace{0.1cm}
\begin{center}
\includegraphics[width=0.8\textwidth]{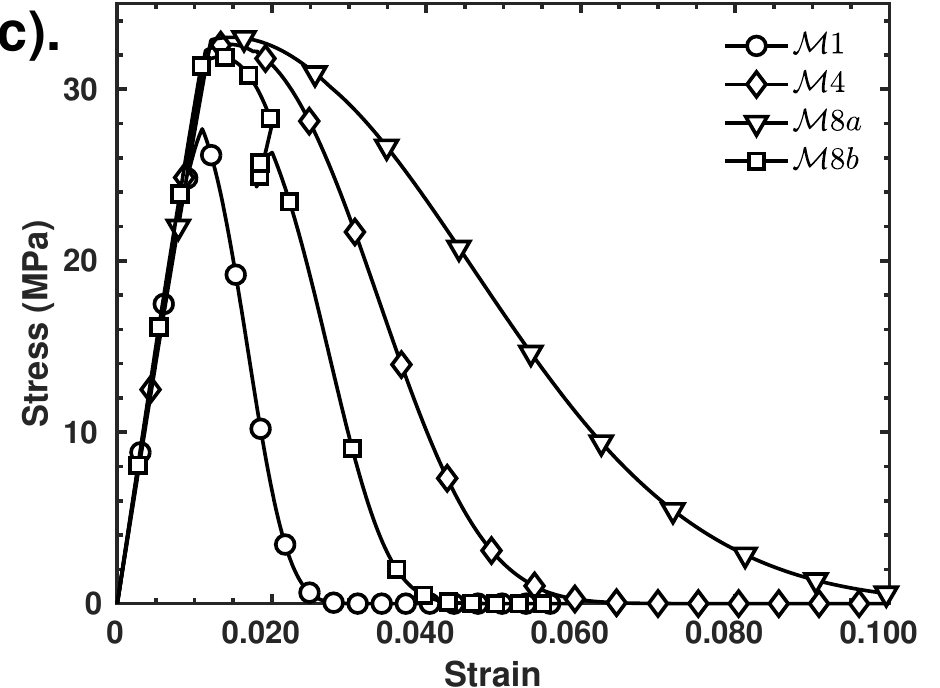}
\par\end{center}%
\end{minipage}
\end{figure}

Based on this calculated $\mathcal{L}_{c}$ the stress strain law
for each partition is modified by adjusting the value of '$m$' for
each partition. Fig. shows stress-strain laws for three cases; Fig.
\ref{fig:stress-strain}(a) represents the variation of stress with
respect to strain when the same characteristic length is used for
$\mathcal{M}1$, $\mathcal{M}4$ and $\mathcal{M}8$; Fig. \ref{fig:stress-strain}(b)
shows the stress-strain variation for four sizes of partitions in
the absence of plastic strains for different characteristic lengths;
Fig. \ref{fig:stress-strain}(c) shows the stress variation when the
plastic strains are taken into account. 

\subsection{Calculation of eigen strain}

The eigen strain is calculated corresponding to macroscopic strain
which is applied incrementally at material/integration point. Loading
at each integration point is equivalent to applied load on the RVE.
Suppose state of a integration point at $n^{th}$ load step is defined
by macroscopic strain $^{n}\boldsymbol{\bar{\epsilon}}$, eigen strain
$^{n}\boldsymbol{\bar{\mu}}^{(\alpha)}$, damage state $^{n}\omega^{(\alpha)}$
and plastic strain $^{n}\boldsymbol{\epsilon}_{p}^{(\alpha)}$. For
a macroscopic strain increment $^{n}\boldsymbol{\triangle\bar{\epsilon}}$,
using current state data $(n+1)^{th}$ load step state variables are
calculated using following procedure:
\begin{enumerate}
\item Using $\Delta\boldsymbol{\bar{\epsilon}}$, phase strain in the $n^{th}$
load step is calculated as 
\begin{equation}
^{n+1}\Delta\boldsymbol{\epsilon}^{(\beta)}=\mathbf{E}:\left(\tensor*[^{n+1}]{\Delta\boldsymbol{\bar{\epsilon}}}{}\right)+\sum_{\alpha}\mathbf{S}^{(\alpha\beta)}:\left(\tensor*[^{n}]{\left(\Delta\boldsymbol{\mu}^{(\alpha)}\right)}{}\right)
\end{equation}
\item On the basis of $^{n+1}\Delta\boldsymbol{\epsilon}^{(\beta)}$ the
absolute strains $^{n+1}\boldsymbol{\epsilon}_{trial}^{(\beta)}$
is updated for each partition as 
\begin{equation}
^{n+1}\boldsymbol{\epsilon}_{trial}^{(\beta)}=\tensor*[^{n}]{(\boldsymbol{\epsilon}_{e}^{(\beta)})}{}+\tensor*[^{n+1}]{(\Delta\boldsymbol{\epsilon}^{(\beta)})}{}
\end{equation}
\item Based on $^{n+1}\boldsymbol{\epsilon}_{trial}^{(\beta)}$, elastic
strain $^{n+1}\boldsymbol{\epsilon}_{e}^{(\beta)}$ and plastic strain
$^{n+1}\boldsymbol{\epsilon}_{p}^{(\beta)}$ components for each phase
are calculated by using Eq. (\ref{eq:48-1}), (\ref{eq:53-1-1}) and
(\ref{eq:55-2}).
\item Then the damage variable $^{n+1}\omega^{(\beta)}$ is calculated using
updated value of $^{n+1}\kappa_{D}^{(\beta)}$ as 
\begin{equation}
^{n+1}\omega^{(\beta)}=\max\left[^{n}\omega^{(\beta)},\ 1-exp\left(1-\frac{1}{m^{(\beta)}}\left(\frac{^{n+1}\kappa_{D}^{(\beta)}}{\kappa_{Df}^{(\beta)}}\right)\right)^{m^{(\beta)}}\right]
\end{equation}
\item Then the tangent modulus for each partition is calculated as following
by using Eq. (\ref{eq:44-1-1}) and (\ref{eq:55-1}).
\item After calculating tangent modulus, eigen strain rate is calculated
at each phase for $(n+1)^{th}$ load step as per following: 
\begin{equation}
\tensor*[^{n+1}]{\left(\frac{\partial\Delta\boldsymbol{\mu}^{(\beta)}}{\partial\Delta\boldsymbol{\epsilon}^{(\beta)}}\right)}{}=\mathbf{I}-\left(\mathbf{L}^{(\beta)}\right)^{-1}:\tensor*[^{n+1}]{\left(\frac{\partial\Delta\boldsymbol{\sigma}^{(\beta)}}{\partial\Delta\boldsymbol{\epsilon}^{(\beta)}}\right)}{}\label{eq:73}
\end{equation}
\item Then strain for each partition is calculated by solving the following:
{\scriptsize{}
\begin{equation}
\left[\begin{array}{c}
\begin{array}{c}
^{n+1}\Delta\boldsymbol{\epsilon}^{(1)}\\
^{n+1}\Delta\boldsymbol{\epsilon}^{(2)}\\
\vdots\\
^{n+1}\Delta\boldsymbol{\epsilon}^{(\beta)}
\end{array}\end{array}\right]=\left[\begin{array}{cccc}
\mathbf{I}-\mathbf{S}^{(11)}\tensor*[^{n+1}]{\left(\frac{\partial\Delta\boldsymbol{\mu}^{(1)}}{\partial\Delta\boldsymbol{\epsilon}^{(1)}}\right)}{} & -\mathbf{S}^{(12)}\tensor*[^{n+1}]{\left(\frac{\partial\Delta\boldsymbol{\mu}^{(2)}}{\partial\Delta\boldsymbol{\epsilon}^{(2)}}\right)}{} & \cdots & -\mathbf{S}^{(1\beta)}\tensor*[^{n+1}]{\left(\frac{\partial\Delta\boldsymbol{\mu}^{(\beta)}}{\partial\Delta\boldsymbol{\epsilon}^{(\beta)}}\right)}{}\\
-\mathbf{S}^{(21)}\tensor*[^{n+1}]{\left(\frac{\partial\Delta\boldsymbol{\mu}^{(1)}}{\partial\Delta\boldsymbol{\epsilon}^{(1)}}\right)}{} & \mathbf{I}-\mathbf{S}^{(22)}\tensor*[^{n+1}]{\left(\frac{\partial\Delta\boldsymbol{\mu}^{(2)}}{\partial\Delta\boldsymbol{\epsilon}^{(2)}}\right)}{} & \cdots & -\mathbf{S}^{(2\beta)}\tensor*[^{n+1}]{\left(\frac{\partial\Delta\boldsymbol{\mu}^{(\beta)}}{\partial\Delta\boldsymbol{\epsilon}^{(\beta)}}\right)}{}\\
\vdots & \vdots & \ddots & \vdots\\
-\mathbf{S}^{(\alpha1)}\tensor*[^{n+1}]{\left(\frac{\partial\Delta\boldsymbol{\mu}^{(1)}}{\partial\Delta\boldsymbol{\epsilon}^{(1)}}\right)}{} & -\mathbf{S}^{(\alpha2)}\tensor*[^{n+1}]{\left(\frac{\partial\Delta\boldsymbol{\mu}^{(2)}}{\partial\Delta\boldsymbol{\epsilon}^{(2)}}\right)}{} & \cdots & \mathbf{I}-\mathbf{S}^{(\alpha\beta)}\tensor*[^{n+1}]{\left(\frac{\partial\Delta\boldsymbol{\mu}^{(\beta)}}{\partial\Delta\boldsymbol{\epsilon}^{(\beta)}}\right)}{}
\end{array}\right]^{-1}\left[\begin{array}{c}
\begin{array}{c}
\mathbf{E}^{(1)}\tensor*[^{n+1}]{\left(\Delta\boldsymbol{\bar{\epsilon}}\right)}{}\\
\mathbf{E}^{(2)}\tensor*[^{n+1}]{\left(\Delta\boldsymbol{\bar{\epsilon}}\right)}{}\\
\vdots\\
\mathbf{E}^{(\beta)}\tensor*[^{n+1}]{\left(\Delta\boldsymbol{\bar{\epsilon}}\right)}{}
\end{array}\end{array}\right]
\end{equation}
}{\scriptsize \par}
\item Now the eigen strain increment is calculated using eq. (\ref{eq:73})
as 
\begin{equation}
\tensor*[^{n+1}]{\left(\Delta\boldsymbol{\mu}^{(\beta)}\right)}{}=\tensor*[^{n+1}]{\left(\frac{\partial\Delta\boldsymbol{\mu}^{(\beta)}}{\partial\Delta\boldsymbol{\epsilon}^{(\beta)}}\right)}{}\tensor*[^{n+1}]{\left(\Delta\boldsymbol{\epsilon}^{(\beta)}\right)}{}
\end{equation}
\end{enumerate}

\subsection{Simulation results}

\begin{figure}[H]
\begin{centering}
\begin{minipage}[t]{0.485\textwidth}%
\begin{center}
\hspace*{0mm}\includegraphics[width=1\textwidth]{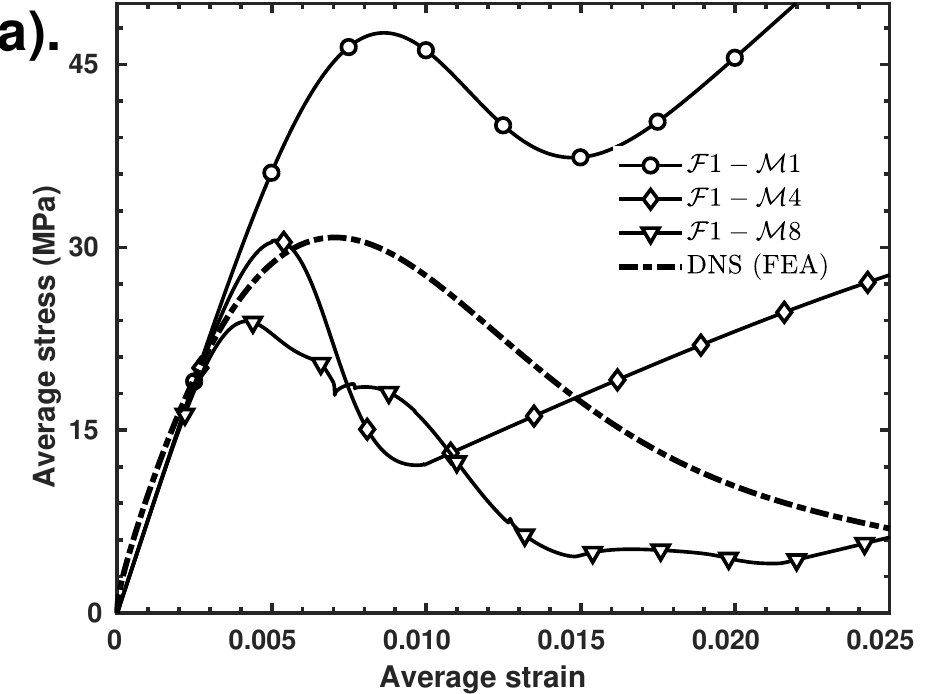}\\[-1.65cm]\hspace*{-30mm}\includegraphics[width=1.00cm]{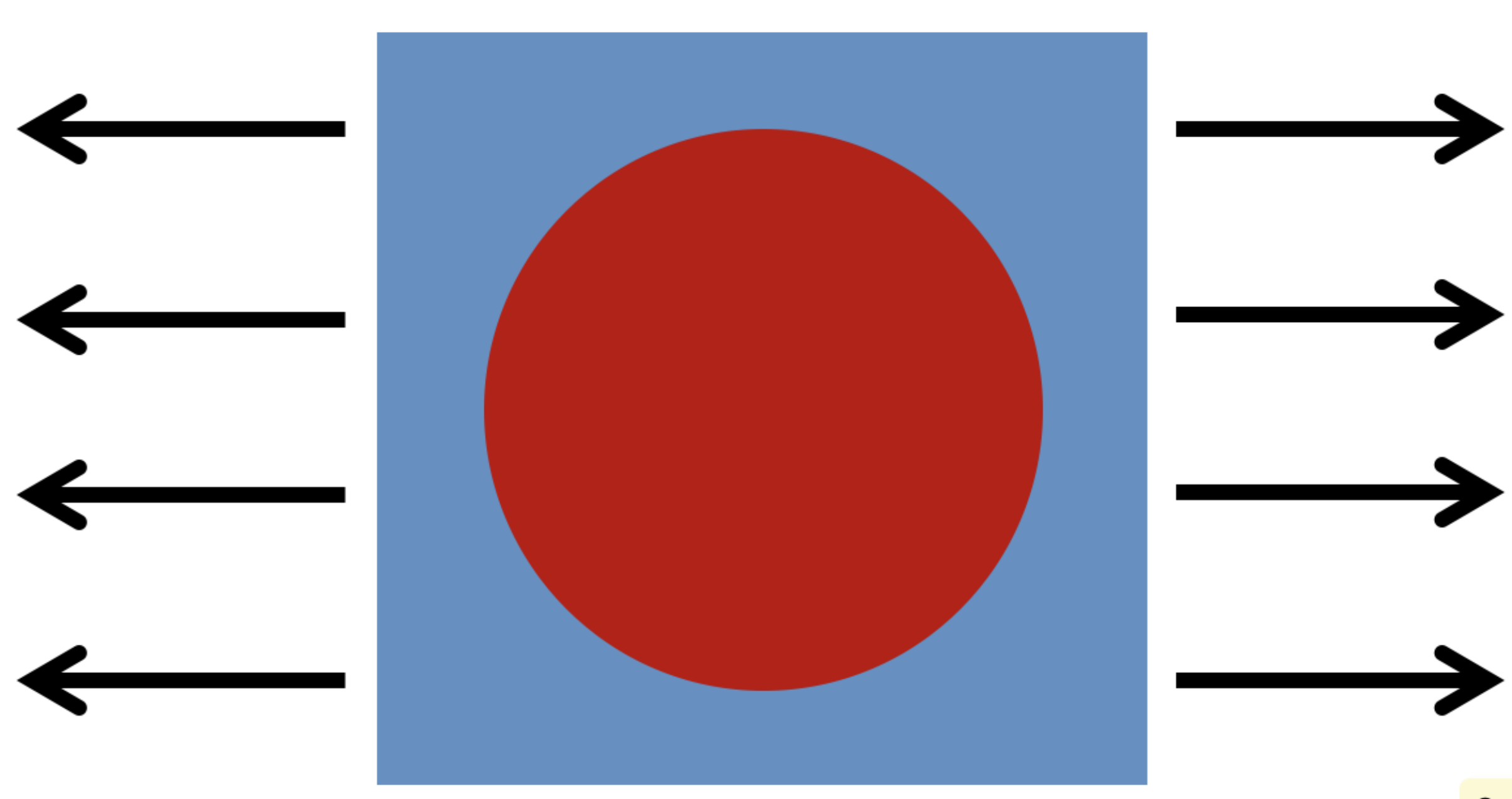}
\par\end{center}%
\end{minipage}
\begin{minipage}[t]{0.485\textwidth}%
\begin{center}
\hspace*{0mm}\includegraphics[width=1\textwidth]{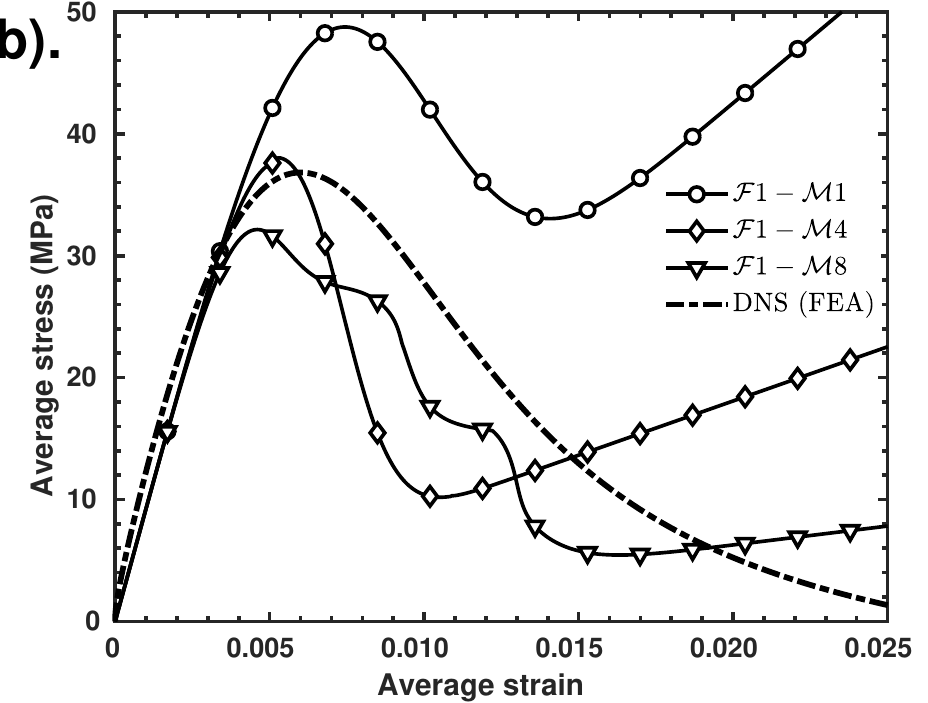}\\[-1.95cm]\hspace*{-30mm}\includegraphics[width=1.0cm]{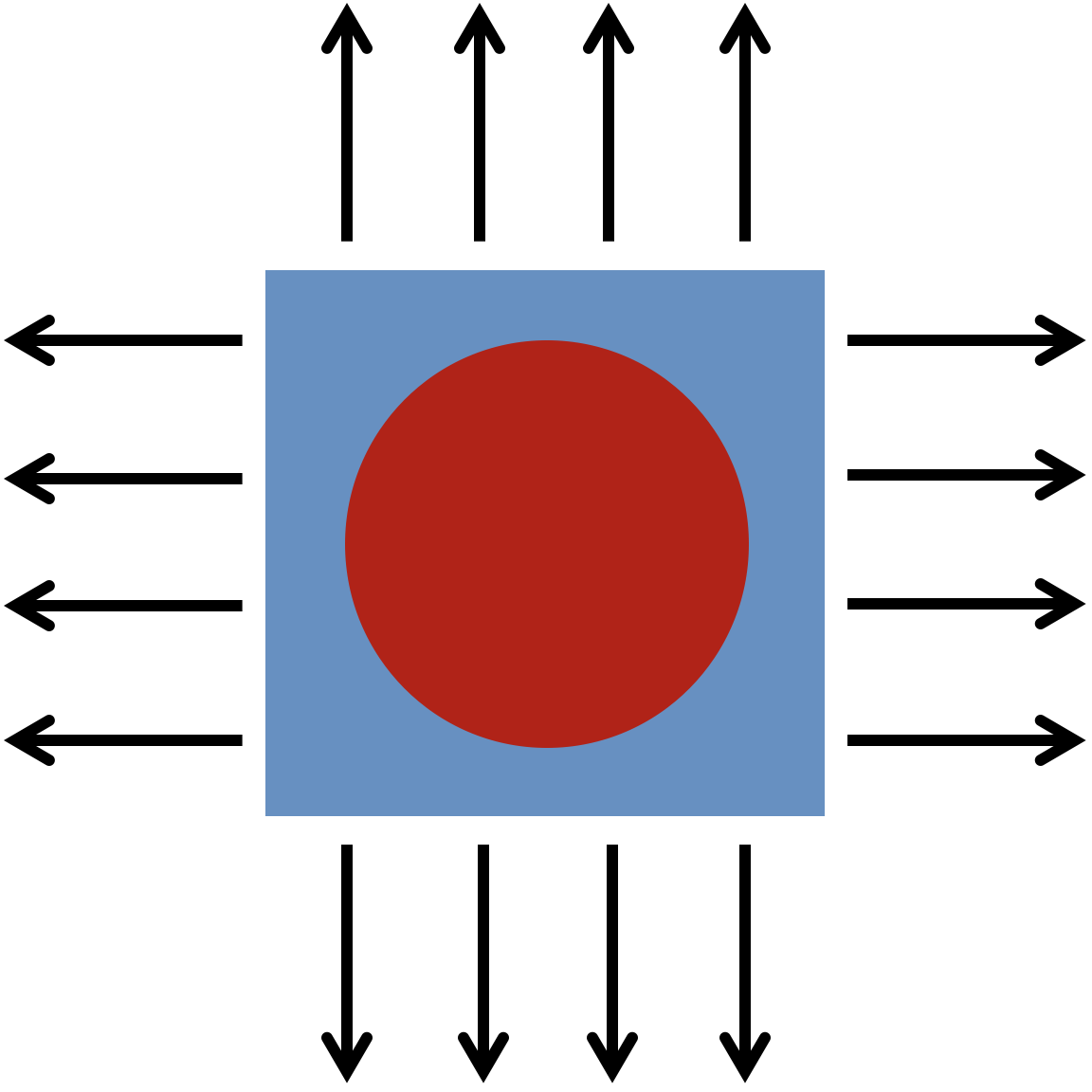}
\par\end{center}%
\end{minipage}
\par\end{centering}
\centering{}\caption{Stress response of unit cell subjected to (a). uniaxial and (b). biaxial
strain without consideration of characteristic length. Unit cell domain
is considered as elastic.\label{fig:default-results}}
\end{figure}

The proposed AEH formulation is checked by performing uniaxial and
biaxial loading simulations for a unit cell. These simulations demonstrate
the effect of increasing the number of partitions and characteristic
length associated with each partition. The influence functions and
coefficient tensors, which define AEH model, are computed as preprocessing
step. For the bench-marking comparison, the results of each AEH based
simulation is compared with the direct numerical simulation (DNS)
using Finite Element Analysis (FEA) of the unit cell. The initial
study is performed by ignoring the effect of characteristic length.
In that case same stress-strain law is used for each matrix partition
as shown in Fig. \ref{fig:stress-strain}(a). Fig. \ref{fig:default-results}
shows the results for three cases $\mathcal{F}1-\mathcal{M}1$, $\mathcal{F}1-\mathcal{M}4$
and $\mathcal{F}1-\mathcal{M}8$ and their comparison with DNS results.
Significant mismatch between AEH predictions and the results of DNS
has been found. Softening starts early as the number of partitions
increase where as post softening behavior shows no trend for both
uniaxial and biaxial loading conditions.

\begin{figure}[H]
\centering{}%
\begin{minipage}[t]{0.485\textwidth}%
\begin{center}
\hspace*{0mm}\includegraphics[width=1\textwidth]{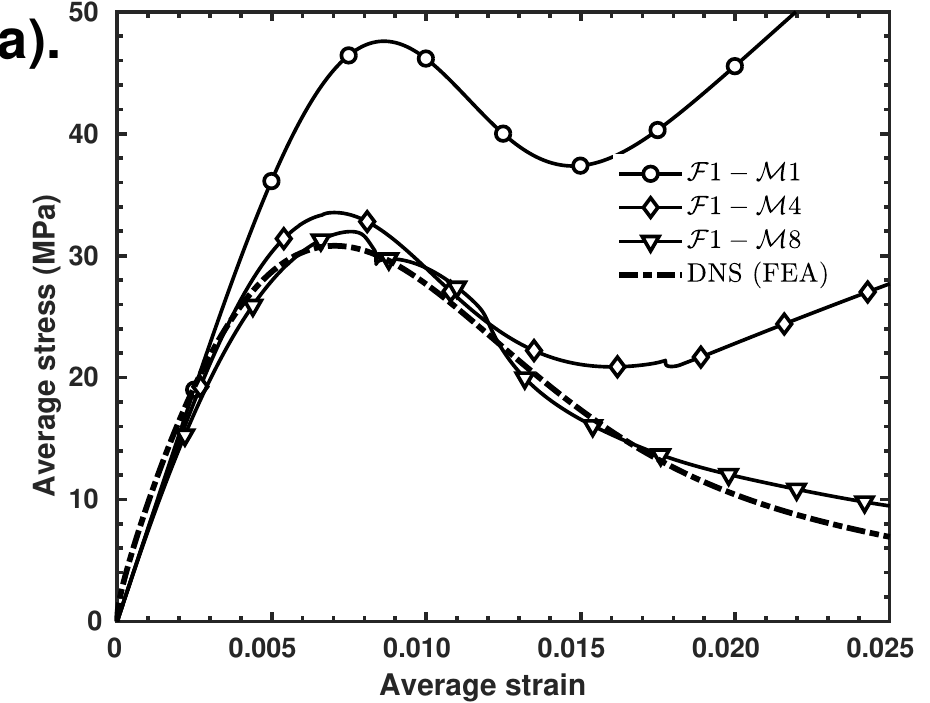}\\[-1.65cm]\hspace*{-30mm}\includegraphics[width=1.0cm]{Uniaxial}
\par\end{center}%
\end{minipage}
\begin{minipage}[t]{0.485\textwidth}%
\begin{center}
\hspace*{0mm}\includegraphics[width=1\textwidth]{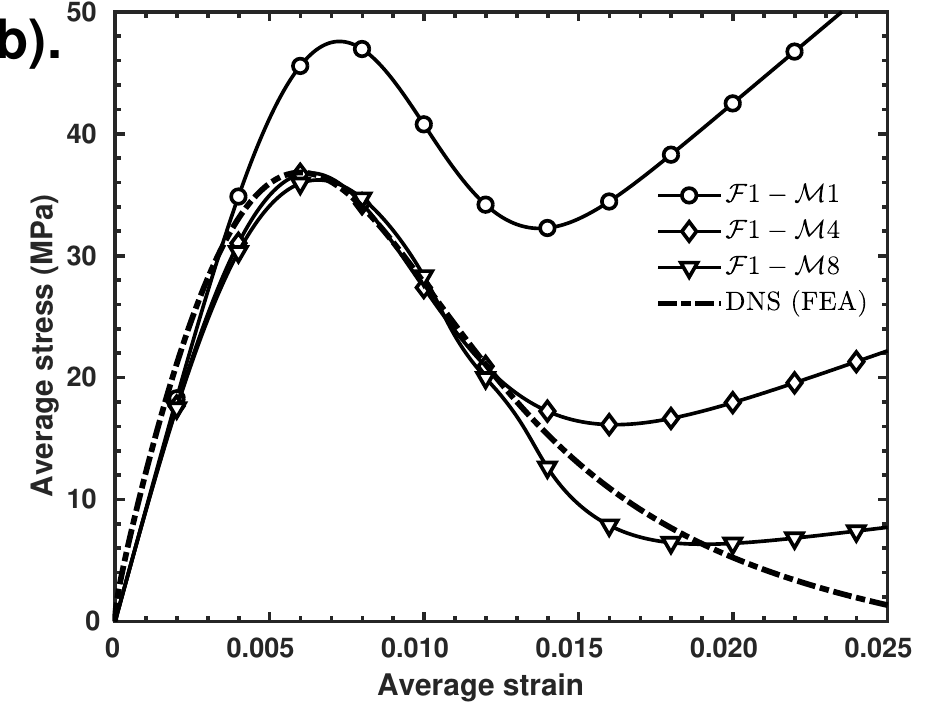}\\[-1.95cm]\hspace*{-30mm}\includegraphics[width=1.0cm]{Biaxial}
\par\end{center}%
\end{minipage}\caption{Stress response of unit cell under subjected to (a). uniaxial and
(b). biaxial strain with consideration of characteristic length. Unit
cell domain is considered as elastic.\label{fig:damage-results}}
\end{figure}

\begin{figure}[H]
\centering{}%
\begin{minipage}[t]{0.485\textwidth}%
\begin{center}
\hspace*{0mm}\includegraphics[width=1\textwidth]{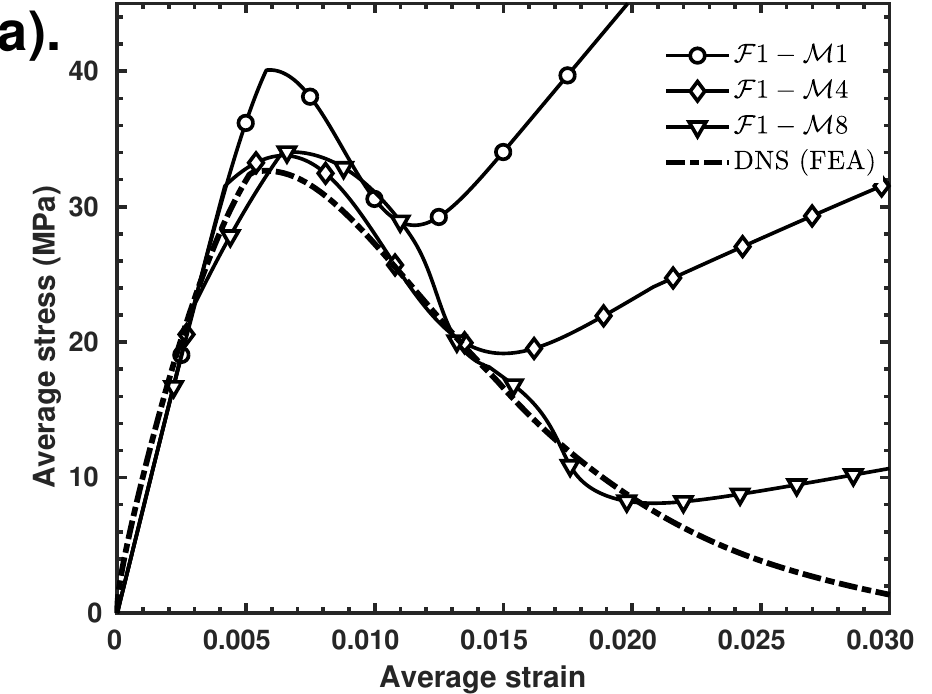}\\[-1.65cm]\hspace*{-30mm}\includegraphics[width=1.0cm]{Uniaxial}
\par\end{center}%
\end{minipage}
\begin{minipage}[t]{0.485\textwidth}%
\begin{center}
\hspace*{0mm}\includegraphics[width=1\textwidth]{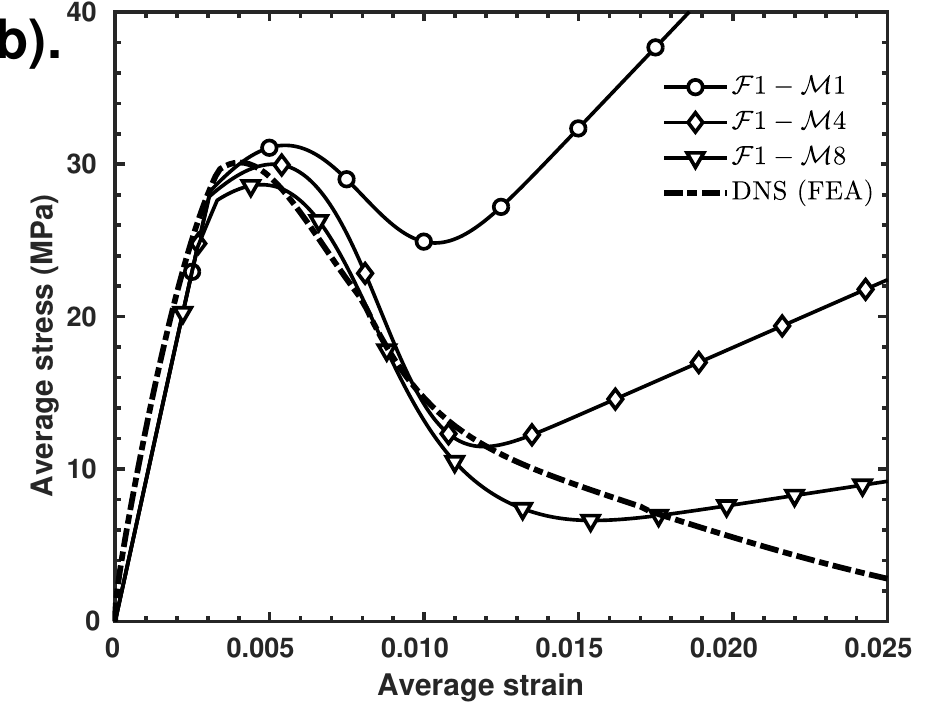}\\[-1.95cm]\hspace*{-30mm}\includegraphics[width=1.0cm]{Biaxial}
\par\end{center}%
\end{minipage}\caption{Stress response of unit cell under subjected to (a). uniaxial and
(b). biaxial strain with consideration of characteristic length. Unit
cell domain is considered as elasto-plastic.\label{fig:plastic-results}}
\end{figure}

Similar simulations are performed by considering the characteristic
length and different stress-strain relationships as shown in Fig.
\ref{fig:stress-strain}(b) for different partitions. Fig. \ref{fig:damage-results}
shows the results for those simulations when whole domain is regarded
as elastic. Similar mismatch between DNS and $\mathcal{F}1-\mathcal{M}1$
results are found due to absence of characteristic length effects.
This also raises the necessity of careful choice of unit cell for
single matrix and single fiber partition model. The results for $\mathcal{F}1-\mathcal{M}4$
match closely with DNS as the number of matrix partitions increases.
Pseudo stiffness effects in the post softening regime are seen for
$\mathcal{F}1-\mathcal{M}4$ model which reduces subsequently with
further increase in number of matrix partitions as for $\mathcal{F}1-\mathcal{M}8$
model. Considerable match between $\mathcal{F}1-\mathcal{M}8$ and
DNS result is found for both uniaxial and biaxial loading cases in
softening regime. Further the proposed formulation is validated for
the unit cell by considering the plastic strain in the matrix region.
Similar match between DNS and homogenization results are found for
all the three partition models when subjected to uniaxual and biaxial
tensile loads as shown in Fig. \ref{fig:plastic-results}. Again the
results from $\mathcal{F}1-\mathcal{M}8$ model match closely with
DNS than others. 

\section{Macroscale Simulation\label{sec:Macoscale-Simulation}}

\begin{figure}[H]
\centering{}\includegraphics[width=1\textwidth]{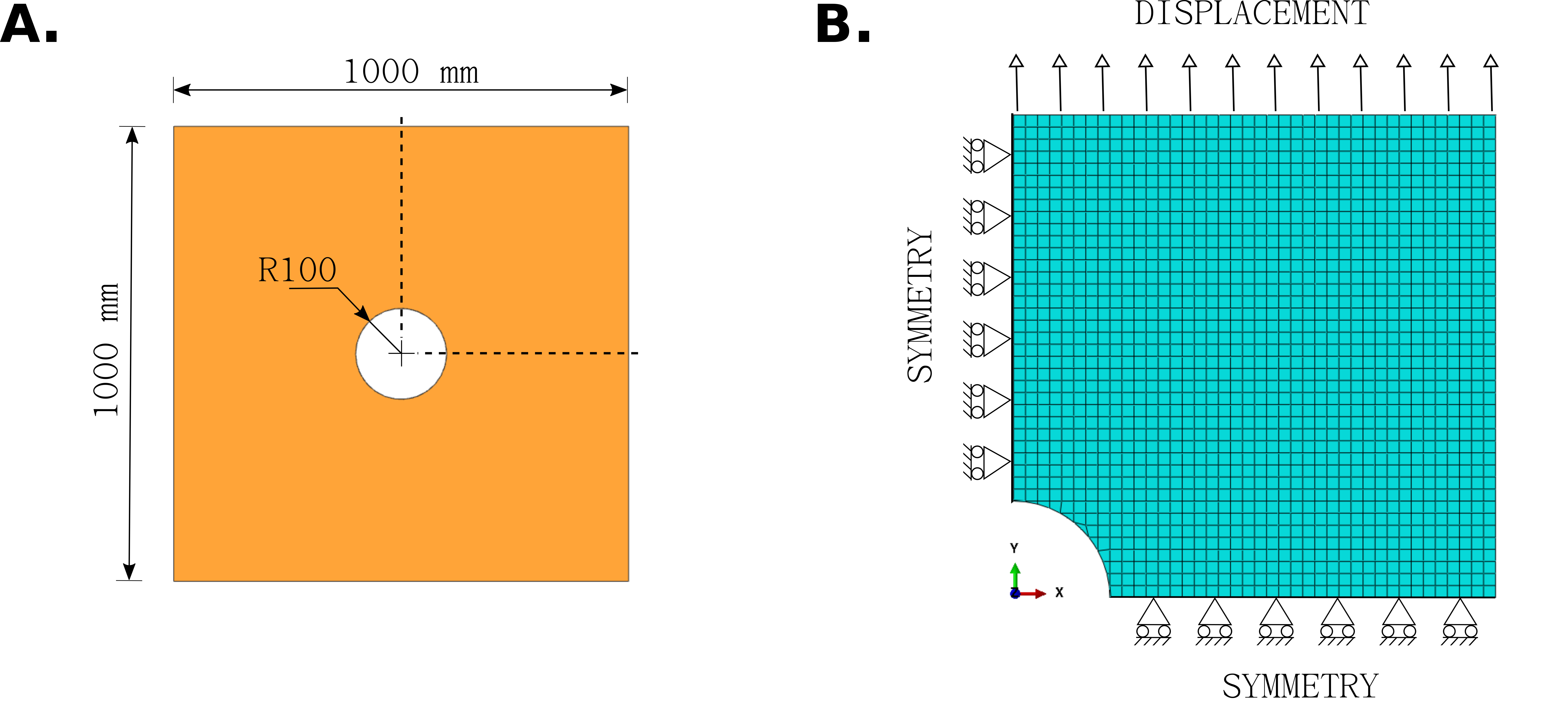}\caption{Macroscale model descriptions (A). Plate with hole showing dimensional
details and symmetry lines (marked dotted) for 1/4 model and (B).
Quarter analysis model with applied boundary \& loading conditions.
\label{fig:FE model}}
\end{figure}

To further demonstrate the capabilities of the proposed AEH based
reduced order model a full two scale analysis is performed which is
verified with direct numerical simulation model. Both DNS and AEH
analyses are carried out using ABAQUS. The user material subroutine
is created for AEH simulation for updating the stress corresponding
to each strain increment. The influence functions and coefficient
tensors, used for AEH simulation, are pre-calculated by performing
the RVE simulations and using the data resulted from these simulations
in the form of ABAQUS output database ({*}.odb) files. Python scripts
are used for the extraction of the data and calculation of the coefficient
tensors. A plate of size $1000$ mm $\times$$1000$ mm with a circular
opening at center of size R$100$ mm as illustrated in Fig. \ref{fig:FE model}(a)
is subjected to uniaxial tensile load. The simulation model is $1/4$
model by considering the geometrical and loading symmetry.

\vspace{1cm}

The whole domain is discretized into $12.5$ mm $\times$$12.5$ mm
unit cells with fiber of $\phi10$ mm embedded at the center of each
unit cell. DNS model is meshed with quadratic elements of 1 mm size
where as for homogenized domain each unit cell is meshed with single
quadratic element.

\begin{figure}[H]
\centering{}%
\begin{minipage}[t]{0.48\textwidth}%
\begin{center}
\includegraphics[width=0.95\textwidth]{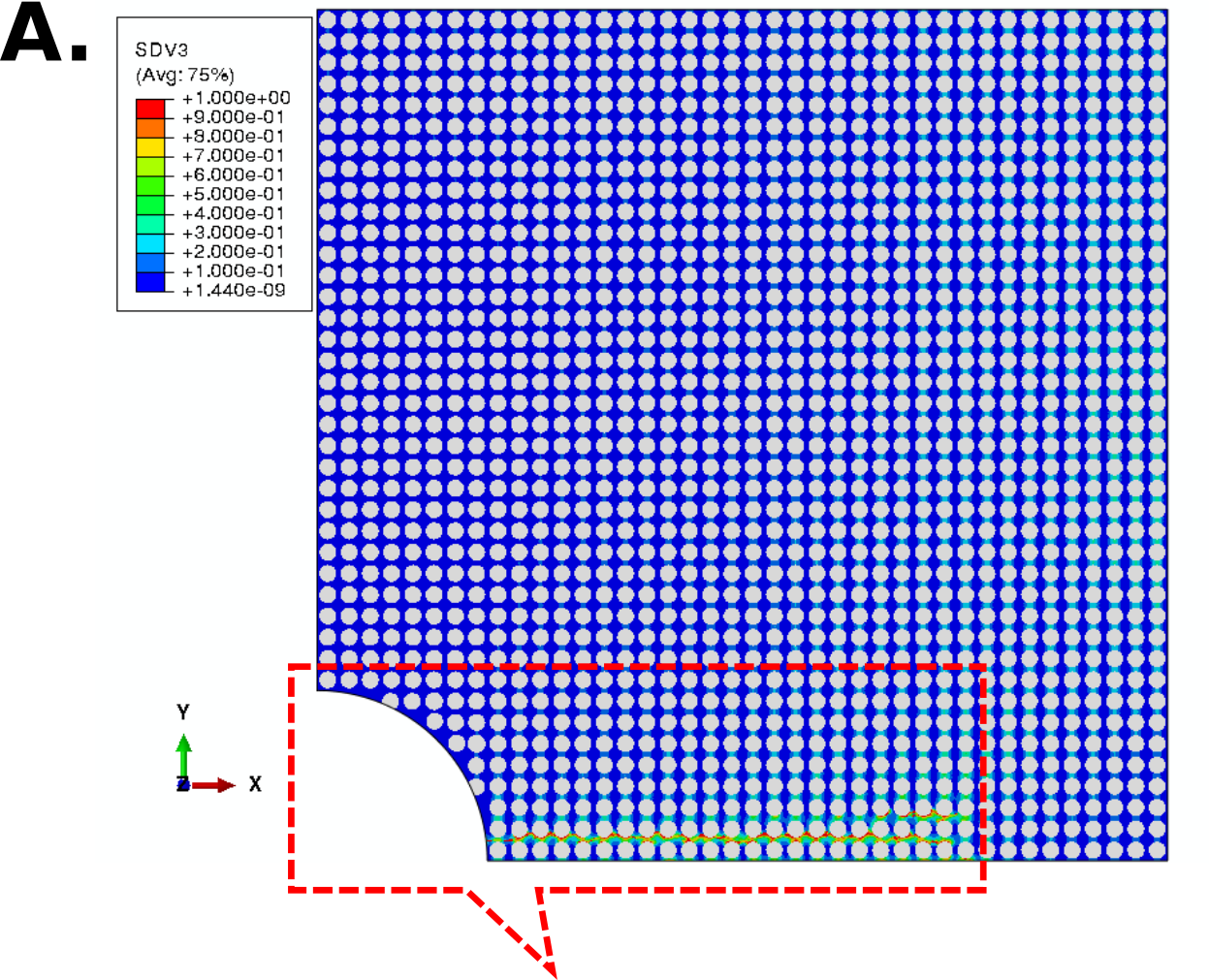}\vspace{0.5mm}
\includegraphics[scale=0.37]{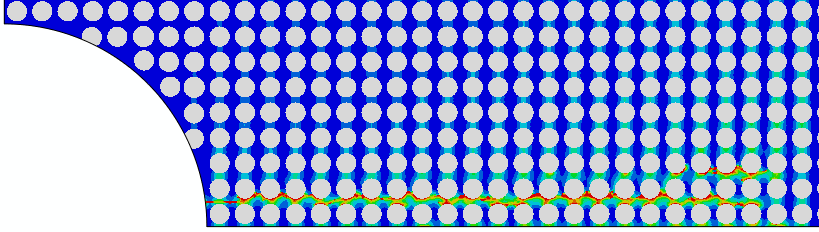}
\par\end{center}%
\end{minipage}\hspace{0.2cm}%
\begin{minipage}[t]{0.48\textwidth}%
\begin{center}
\includegraphics[width=0.98\textwidth]{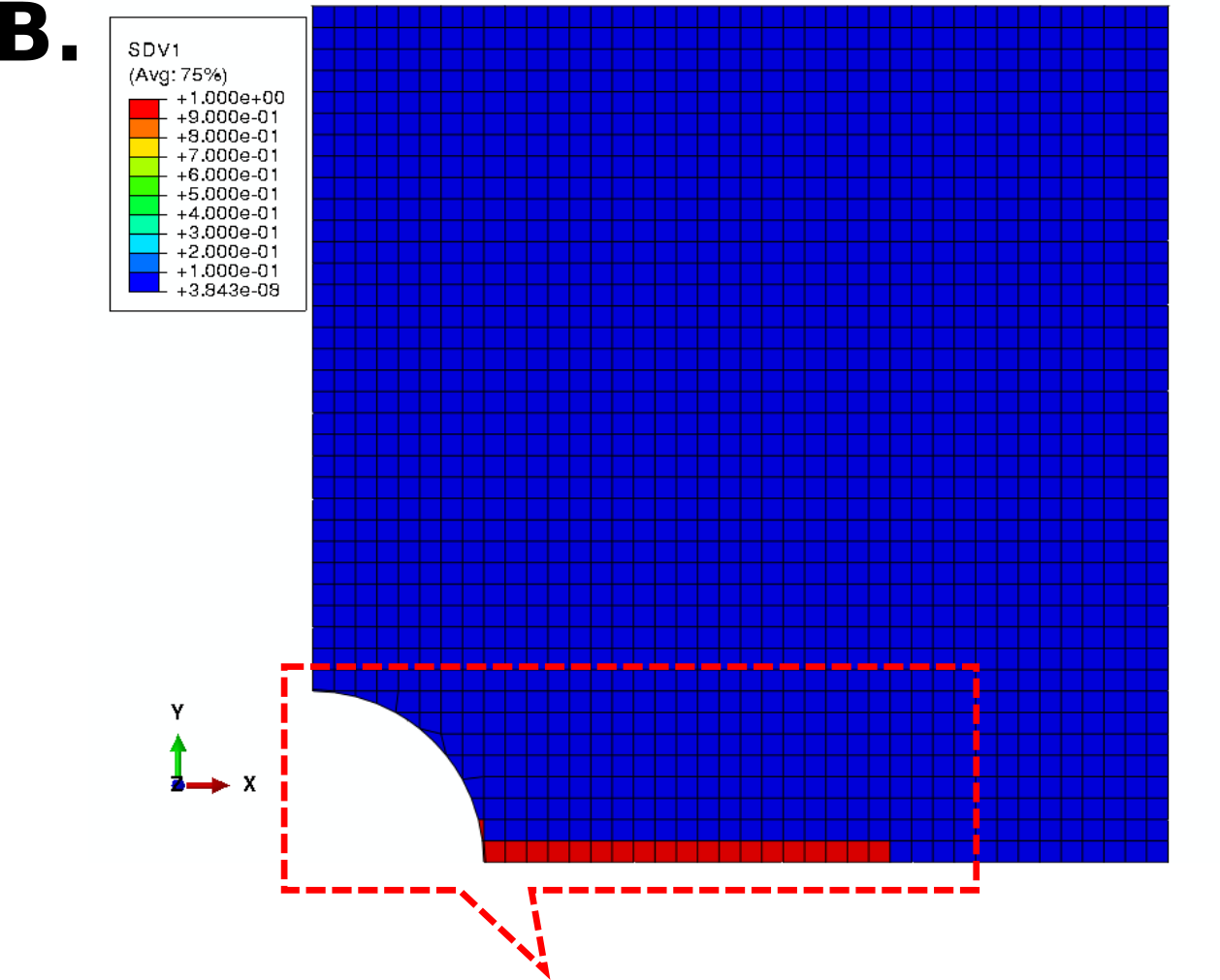}\vspace{0.5mm}
\includegraphics[scale=0.44]{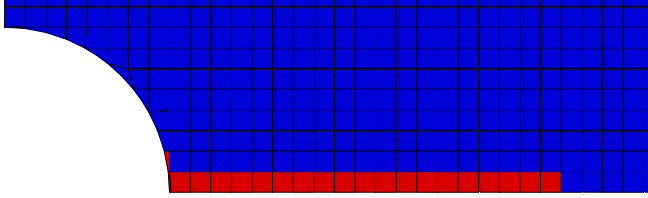}
\par\end{center}%
\end{minipage}\caption{Comparison of damage map obtained from the simulation of plate made
of (A). Heterogeneous media and (B). Equivalent homogenized media
subjected to uniaxial tension. \label{fig:damage-map}}
\end{figure}

\begin{figure}[H]
\centering{}%
\begin{minipage}[t]{0.48\textwidth}%
\begin{center}
\includegraphics[width=0.95\textwidth]{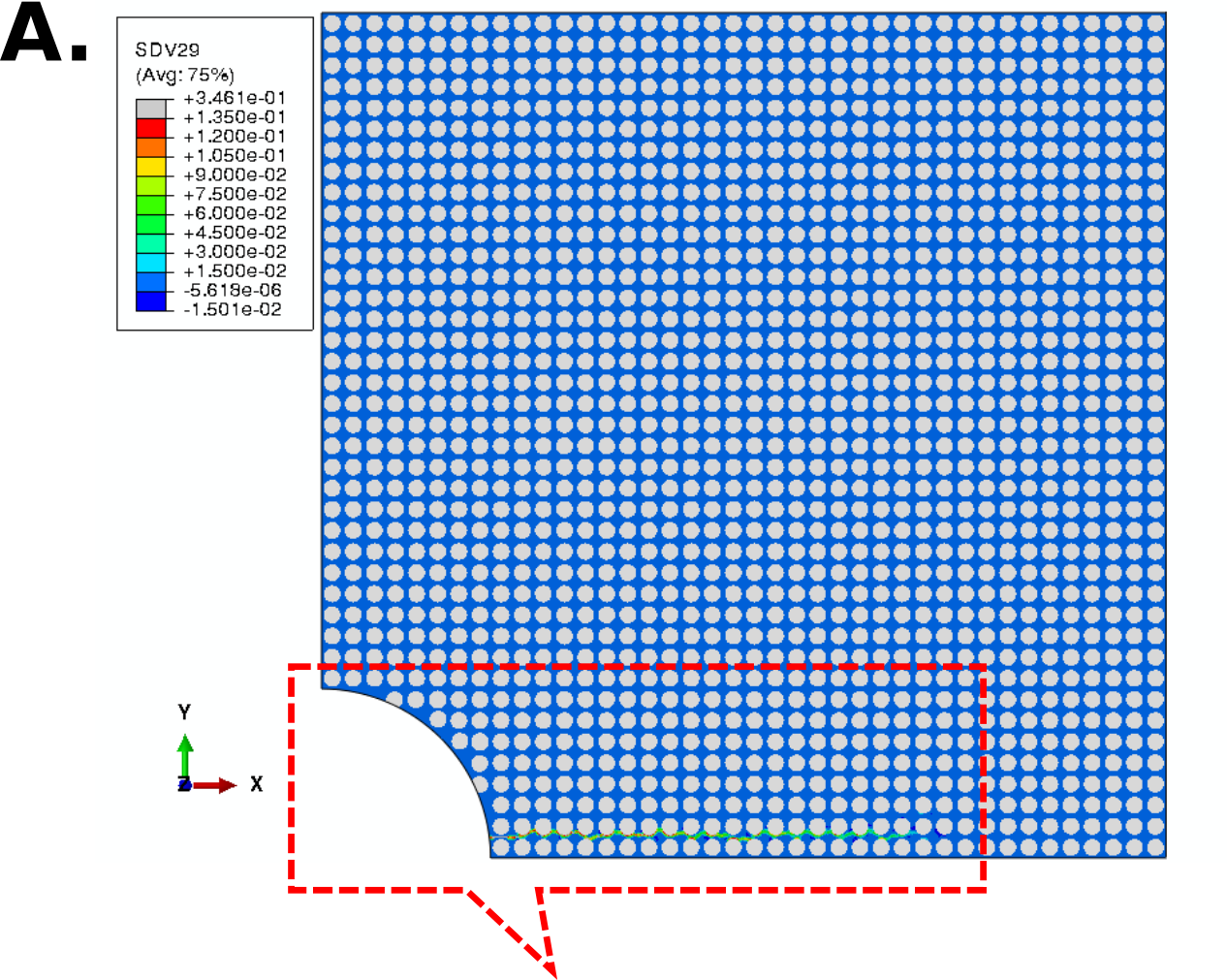}\vspace{0.5mm}
\includegraphics[scale=0.42]{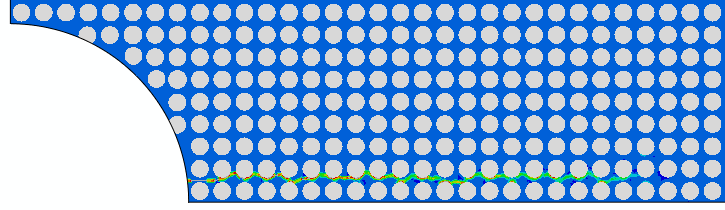}
\par\end{center}%
\end{minipage}\hspace{0.2cm}%
\begin{minipage}[t]{0.48\textwidth}%
\begin{center}
\includegraphics[width=0.95\textwidth]{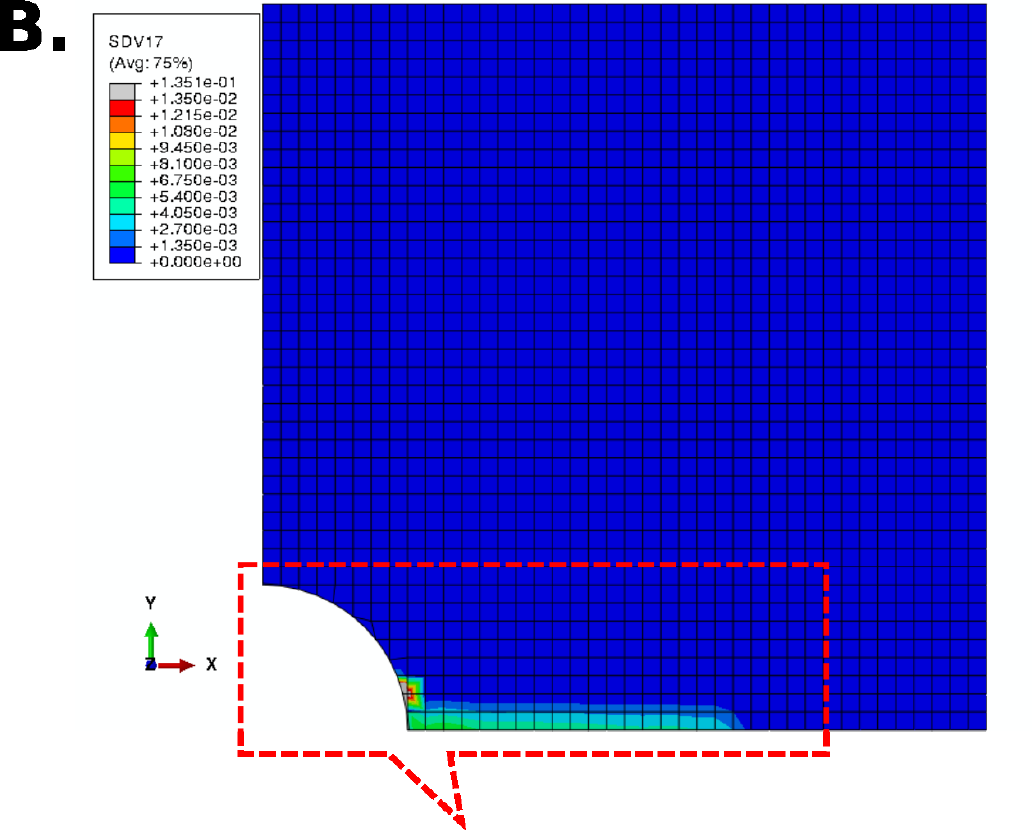}\vspace{0.5mm}
\includegraphics[scale=0.44]{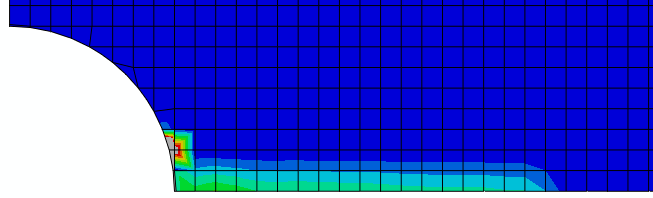}
\par\end{center}%
\end{minipage}\caption{Comparison of equivalent plastic strain obtained from the simulation
of plate made of (A). Heterogeneous media and (B). Equivalent homogenized
media subjected to uniaxial tension. \label{fig:plastic-map}}
\end{figure}

\textcolor{black}{{} It is evident that AEH approach
efficiently reduces the computational complexity of the analysis from
a FE model comprised of 326,912 nodes and 327,892 second order elements
to a system with 3,276 nodes and 1,557 elements, effectively 653,824
degrees of freedom to 22,932. Both the simulations are performed on
a LINUX Workstation HP Z820 having Intel Xeon processor 3.5 GHz with
12 core, 128 Gb RAM and Quadro K4000 NVIDIA GPU. The distributed memory
allocation is utilized for each simulation. Tab. 1 shows the wall
time spent for both the simulations which also demonstrates the significant
reduction in the time using AEH based approach. This wall time excludes
the time of AEH preprocessing and reports only running time of macroscale
simulation.}

\begin{table}[H]
\begin{centering}
\textsf{\textbf{\textcolor{black}{Table 1: }}}\textsf{\textcolor{black}{Simulation
mesh details and comparison of wall clock time for DNS and AEH model\bigskip{}
\label{Table-A.1:-Simulation}}}
\par\end{centering}
\centering{}\textcolor{black}{}%
\begin{tabular}{>{\centering}p{5cm}>{\centering}p{5cm}>{\centering}p{4cm}}
\hline 
\noalign{\vskip0.1cm}
\textbf{\textcolor{black}{Parameters}} & \textbf{\textcolor{black}{DNS Model}} & \textbf{\textcolor{black}{AEH Model}}\tabularnewline[0.1cm]
\hline 
\noalign{\vskip0.1cm}
\textcolor{black}{No. of elements} & \textcolor{black}{327,892} & \textcolor{black}{1,557}\tabularnewline[0.1cm]
\noalign{\vskip0.1cm}
\textcolor{black}{No. of nodes} & \textcolor{black}{326,912} & \textcolor{black}{3,276}\tabularnewline[0.1cm]
\noalign{\vskip0.1cm}
\textcolor{black}{Total degrees of freedom} & \textcolor{black}{653,824} & \textcolor{black}{22,932}\tabularnewline[0.1cm]
\noalign{\vskip0.1cm}
\textcolor{black}{Wall time (hours:min)} & \textcolor{black}{02:35} & \textcolor{black}{00:17}\tabularnewline[0.1cm]
\hline 
\end{tabular}
\end{table}

\begin{figure}[H]
\centering{}\includegraphics[width=0.5\textwidth]{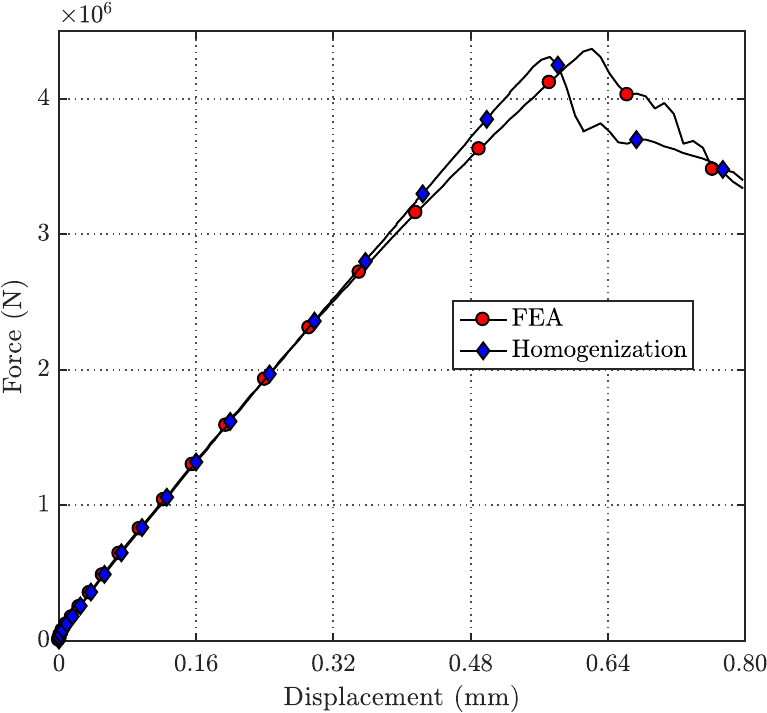}\caption{Comparison of Force-Displacement data. \label{fig:fd-diagram}}
\end{figure}

Under the applied displacement plate deforms inelastically as well
as damage occurs near the highly stressed zone around the hole. Fig.
\ref{fig:damage-map} and Fig. \ref{fig:plastic-map} show the comparison
of damage map and plastic strain from DNS and homogenization model
under the applied displacement. The accuracy of these variables is
further evidenced by force-displacement plot as shown in Fig. \ref{fig:fd-diagram}.

\section{Conclusions}

A reduced order asymptotic expansion homogenization technique for
capturing the plasticity and damage effects in a heterogenous material
is presented in this manuscript. Discretization of RVE may lead to
'local' formulation while capturing the damage in a material. The
localization limiter which were used for AEH approaches in past are
unable to change the problem to 'nonlocal'. A novel methodology is
proposed to alleviate the problem of strain localization which is
caused due to reduction of order of macroscale domain. The problem
of post failure pseudo stiffness is also discussed and an optimal
way of partitioning the microscale domain is presented. The proposed
formulation is validated by comparing the homogenized calculation
data with direct numerical solution of a RVE. Finally a macroscopic
problem is solved for a homogenized domain and results are compared
with FE simulation of heterogenous material.

\newpage
\bibliographystyle{unsrtnat}
\bibliography{mybib}

\end{document}